\documentclass[aps,prb,twocolumn]{revtex4-1}

\usepackage{graphicx}
\usepackage{bm}
\usepackage{amsmath}
\usepackage{amssymb}
\usepackage{euscript}
\usepackage{verbatim}
\usepackage{wrapfig}
\usepackage{setspace}
\usepackage{xcolor}	
\usepackage{amsfonts}
\usepackage{lipsum}
\usepackage[caption=false]{subfig}
\usepackage{flushend}

\usepackage[]{natbib}

\usepackage[
colorlinks=true,
urlcolor=black,
linkcolor=blue,
citecolor=black
]{hyperref}

\newcommand{\Tr}{\mathrm{Tr}}

\newcommand{\ket}[1]{\left| #1 \right\rangle}
\newcommand{\bra}[1]{\left\langle #1 \right|}

\begin{document}
	
	%\preprint{APS/123-QED}
	
	\title{Dynamics of Charge-Resolved Entanglement after a Local Quench}

	\author{Noa Feldman}
	\author{Moshe Goldstein}
	\affiliation{Raymond and Beverly Sackler School of Physics and Astronomy, Tel-Aviv University, 6997801 Tel Aviv, Israel}

	\begin{abstract}
		Quantum entanglement and its main quantitative measures, the entanglement entropy and entanglement negativity, play a central role in many body physics. An interesting twist arises when the system considered has symmetries leading to conserved quantities: Recent studies introduced a way to define, represent in field theory, calculate for 1+1D conformal systems, and measure, the contribution of individual charge sectors to the entanglement measures between different parts of a system in its ground state. In this paper, we apply these ideas to the time evolution of the charge-resolved contributions to the entanglement entropy and negativity after a local quantum quench. 
		We employ conformal field theory techniques and find that the known dependence of the total entanglement on time after a quench, $S_A \sim \log(t)$, results from $\sim\sqrt{\log(t)}$ significant charge sectors, each of which contributes $\sim\sqrt{\log(t)}$ to the entropy.  We compare our calculation to numerical results obtained by the time-dependent density matrix renormalization group algorithm and exact solution in the noninteracting limit, finding good agreement between all these methods. 
	\end{abstract}
	
	%\pacs{Valid PACS appear here}% PACS, the Physics and Astronomy
	% Classification Scheme.
	%\keywords{Suggested keywords}%Use showkeys class option if keyword
	%display desired
	\maketitle
	
	%\tableofcontents
	
	\section{\label{sec:introduction}Introduction}
	The discussion of entanglement started in the early days of quantum mechanics by Einstein, Podolsky, and Rosen~\cite{EPR} as well as Schr\"odinger~\cite{Schrdinger}, yet quantum entanglement remains an active topic of research in several fields of quantum theory ~\cite{Srednicki93,RST06,CC,Nishioka09}, and specifically in quantum many body systems~\cite{Amico08,H4,Laflorencie}. Entanglement in many body systems is used for elucidating their physics~\cite{Osterloh,Osborne,Vidal,CC}, for understanding the limits of simulating quantum systems on a classical computer~\cite{Cirac_2009,Verstraete,SCHOLLWOCK,Orus}, and to characterize their utility as a resource for various quantum information applications~\cite{Ekert,Bennet92,Bennet93,Shor,teleportation,Wootters98,Nielsen,Gisin02,Eisert,Harrow}.

	In order to define the entanglement measures we study in this paper, we introduce the density matrix (DM) for a state $\ket{\psi}$, $\rho = \ket{\psi}\bra{\psi}$. We also define the reduced density matrix (RDM): For two subsystems $A$ and $B$, and a pure state $\ket{\psi}$ of the combined subsystems, the RDM is defined to be $\rho_A = \Tr_B \rho$.
	In this case (pure state of the total system), the basic measure for the entanglement of the subsystem $A$ with its environment $B$ is the \textit{von Neumann entanglement entropy} (vNEE)~\cite{VN}:
	\begin{equation}\label{eq:vNEE}
	S_A = -\Tr \rho_A \log(\rho_A).
	\end{equation}
	We also introduce the moments of the density matrix, which we will refer to as \textit{R\'enyi entropies}: The $n$th R\'enyi entropy (RE) is defined to be:
	\begin{equation}\label{eq:RE}
	S^{(n)}_A = \Tr \rho_A^n.
	\end{equation}
	We stress that the definition of the REs here is \textit{different} from the standard definition, $\ln(\mathrm{Tr}\rho_A^n)/(1-n)$. The REs obey $S_A = -\partial_n S^{(n)}_A |_{n\to 1}$.  The REs are entanglement monotones, but they do not possess all the useful properties that the vNEE does\cite{Wilde}. However, they are easier to calculate, and can be measured experimentally more easily\cite{HorodeckiEkert02,Alves04,Daley,AbaninEugene12,Pichler,Islam,BBB16,Elben18,VEDCZ18,REMeas18,Brydges19,CGSMeas} (although a protocol for the measurement of the spectrum of the RDM of a bosonic system was proposed in Ref.~\onlinecite{Pichler16}).

	%%%
	When the total state of the two considered subsystems is not pure, different entanglement measures are needed. For two subsystems $A_1$ and $A_2$, coupled to an environment $B$, a popular entanglement measure is the \textit{entanglement negativity}~\cite{Peres}:
	\begin{equation}\label{eq:neg}
	\mathcal{N}_{A_1, A_2} = \frac{\left|\left| \rho_{A_1\cup A_2}^{T_2}\right|\right| - 1}{2},
	\end{equation}
	where $||\cdot ||$ denotes the trace norm, and the superscript $T_2$ stands for the partial transpose: 
	\begin{equation*}
	\begin{aligned}
	& \rho_A = \sum_{i, j, i^\prime, j^\prime} c_{i, j, i^\prime, j^\prime} \ket{i}_{A_1}\ket{j}_{A_2}\bra{i^\prime}_{A_1}\bra{j^\prime}_{A_2} \rightarrow \\&
	\rho_A^{T_2} =  \sum_{i, j, i^\prime, j^\prime} c_{i, j, i^\prime, j^\prime} \ket{i}_{A_1}\ket{j^\prime}_{A_2}\bra{i^\prime}_{A_1}\bra{j}_{A_2},
	\end{aligned}
	\end{equation*}
	where $(\ket{i}_{A_1}, \ket{i^\prime}_{A_1})$ and $( \ket{j}_{A_2}, \ket{j^\prime}_{A_2})$ are orthonormal bases for subsystems $A_1$ and $A_2$ respectively.
	Here too we define \textit{R\'enyi negativities} (RNs):
	\begin{equation}\label{eq:RN}
	\mathcal{N}_{A_1,A_2}^{(n)} = \Tr \left(\rho_{A_1\cup A_2}^{T_2}\right)^n,
	\end{equation}
	which could be analytically continued from an even integer $n$ to yield the negativity, using $\left|\left| \rho_{A_1\cup A_2}^{T_2}\right|\right| = \lim_{n\rightarrow 1/2} \mathcal{N}_{A_1,A_2}^{(2n)}$. While the RNs are not even entanglement monotones (as opposed to the REs), they are still useful indicators of entanglement since (like the REs), they are experimentally measurable for both bosons~\cite{CGS,Gray18} and fermions~\cite{CGSMeas}, and are easier to calculate.

	We study a system that has some conserved charge $\hat{N}$, that obeys $\hat{N}_A + \hat{N}_B = \hat{N}$, for $\hat{N}_i$ the charge on subsystem $i$, such as a spin component or particle number. We assume that the state of the total system has some fixed value of $\hat{N}$, hence $\left[\rho, \hat{N}\right] = 0$. Performing a partial trace over the equation above, we get $\left[\rho_A, \hat{N}_A\right] = 0$. This implies that the RDM is block diagonal, each block corresponding to some eigenvalue of $\hat{N}_A$ and denoted by $\rho_A^{(N_A)}$, $\rho_A = \oplus_{N_A}\rho_A^{(N_A)}$. This allows to define the \textit{charge resolved} vNEE and REs~\cite{LR,GS,Xavier18,Barghathi18,Barghathi19}:
	\begin{equation}
	\begin{split}
	& S_A(N_A) = - \Tr \rho_A^{(N_A)} \log \rho_A^{(N_A)}, \\
	& S_A^{(n)}(N_A) = \Tr \left(\rho_A^{(N_A)}\right)^n.
	\end{split}
	\end{equation}
	The blocks $\rho_A^{(N_A)}$ are not normalized and thus $S_A = \sum_{N_A}S_A(N_A), \quad S_A^{(n)} = \sum_{N_A}S_A^{(n)}(N_A)$. A normalized version has been used in some recent works~\cite{EquipS,EquipC,EquipDM}. The corresponding entropies then characterize the entanglement following a projective measurement of the particle number in subsystem $A$. We prefer to follow our earlier convention~\cite{GS, CGS} and \emph{not} to normalize the individual blocks, since the contributions of the unnormalized blocks to the total entanglement are not only more straightforwardly accessible in the calculation, but are also directly accessible experimentally, using either the protocol introduced in our previous works~\cite{GS,CGS} or the method based on random evolution introduced in Refs.~\onlinecite{VEDCZ18,REMeas18}, if number-conserving evolution is used, as in those papers.
	
	For the case of two subsystems coupled to an environment, it is useful to define the charge imbalance: $\hat{Q} = \hat{N}_{A_1} - \hat{N}_{A_2}^{T_2}$. By performing a partial trace over the equation shown above we get~\cite{CGS} $0 = \left([\rho_A, \hat{N}_A]\right)^{T_2} = [\rho_A^{T_2}, \hat{N}_{A_1} - \hat{N}_{A_2}^{T_2}] = [\rho_A^{T_2}, \hat{Q}]$. Note that in a standard Fock space basis, $\hat{N}_{A_2}^{T_2} = \hat{N}_{A_2}$. We may then define the \textit{charge imbalance resolved} negativity and RNs~\cite{CGS}:
	\begin{equation}
	\begin{split}
	& \mathcal{N}_{A_1,A_2}(Q) = \frac{\left|\left|\left(\rho_A^{T_2}\right)^{(Q)}\right|\right| - 1}{2}, \\
	& \mathcal{N}_{A_1,A_2}^{(n)}(Q) = \Tr \left(\left(\rho_A^{T_2}\right)^{(Q)}\right)^n.
	\end{split}
	\end{equation}

	By definition, $S_A^{(1)}(N_A)$ is the charge distribution in subsystem A, and $\mathcal{N}_{A_1,A_2}^{(1)}(Q)$ is the charge imbalance distribution, demonstrating an inherent relation between entanglement and charge distribution. The charge resolved entanglement can be used as an instrument to study entanglement properties and gain a better understanding of the charge block structure of the entanglement spectrum. A particular interest in the relation between charge distribution and entanglement has risen for systems undergoing a local quench: We prepare two subsystems $A$ and $B$ in the ground state, couple the two subsystems at $t=0$ and study the evolution of entanglement between them. 
	In Ref.~\onlinecite{KL} a relation between charge distribution (quantum noise) and the entanglement has been derived for noninteracting fermions following this type of quench, and motivated further study of the relation between entanglement and charge distribution~\cite{Hsu2009, Song2010,Song2011,Song20116, Song2012,Calabrese_2012,Vicari12,Eisler2013,Eisler20138,Chien2014,Petrescu_2014,Thomas2015,Dasenbrook2015,Sinitsyn_2016}. 
	With the charge-resolved entanglement measures just discussed it becomes apparent that one should not separately address the dynamics of charge and entanglement, but rather their combined measures. The goal of this work is to addresses this question.

	\begin{figure*}[ht]
		\subfloat
		[\label{fig:rdm}]{		\includegraphics[width=.3\linewidth]{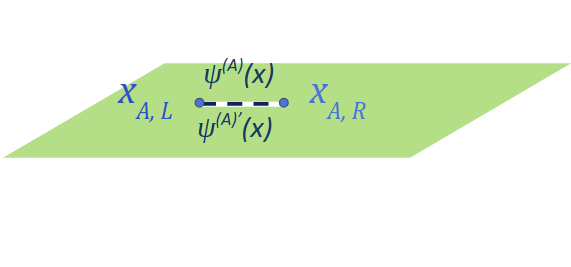}}
		\subfloat
		[\label{fig:sub_replica}]{		\includegraphics[width=.3\linewidth]{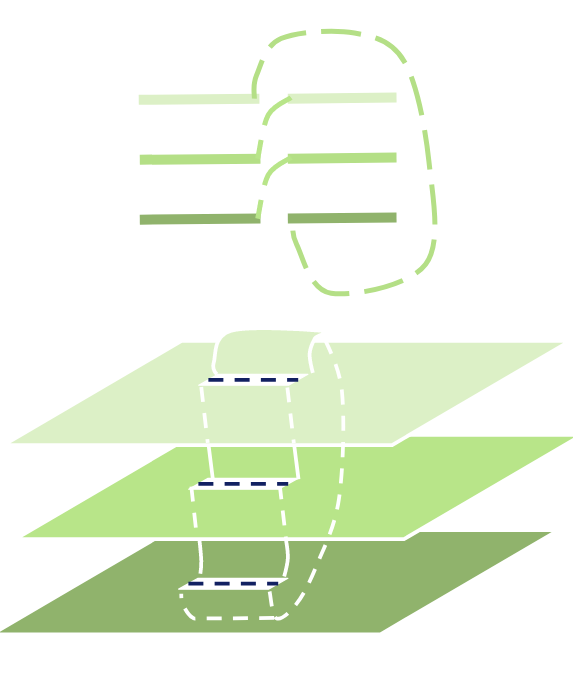} 
		} 
		\subfloat
		[\label{fig:replicaFlux}]{		\includegraphics[width=.3\linewidth]{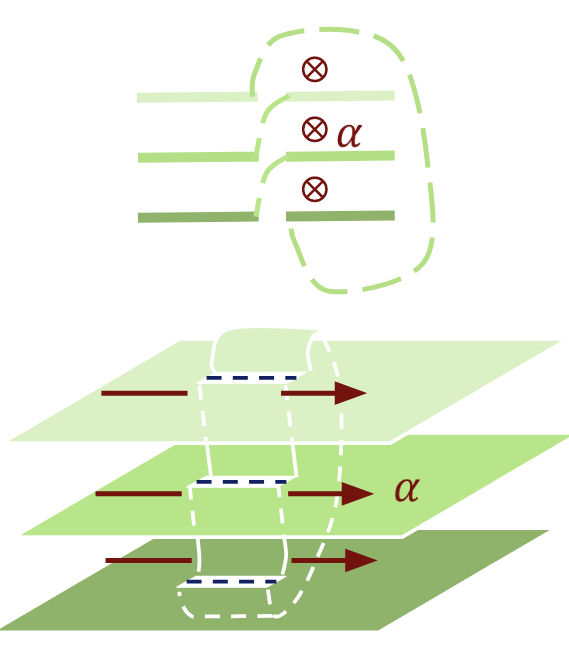}%
		}
		\caption{(a) Geometric representation of the RDM matrix element $\bra{\psi^{(A)}(x)}\rho_A\ket{{\psi^{(A)}}^\prime(x)}$ as an infinite Riemann sheet with a slit corresponding to subsystem $A$, with the indicated boundary conditions on the fields. (b) Side and front view of the geometric representation of the R\'enyi entropy as a path integral over a complex manifold composed of several copies of the RDM. (c) Inserting an AB-flux $\alpha$ coupled to the charge $\hat{N}$ between the Riemann sheets. Particles living on subsystem $A$ will acquire a phase $\alpha$, while particles living on the environment will not be affected by the flux, resulting in a phase $\hat{N}_A \alpha$ acquired by the system.}
		\label{fig:replica}
	\end{figure*}		
	
	In this paper we combine the methods from Refs.~\onlinecite{GS, CGS} for calculating the time dependent charge-resolved entanglement in 1+1D conformal field theory (CFT) systems. We compare them to exact results for the XX model, as well as to time dependent density matrix renormalization group (tDMRG) results for the XXZ model [Eq. (\ref{eq:XXZ})]. For the dynamics of the charge resolved vNEE, we get the contribution of the different charge sectors to the total entropy. We find that the behavior $S_A(t) \sim \log(t)$, as extracted in Ref.~\onlinecite{CC} and decribed below, mostly originates from $\sim \sqrt{\log(t)}$ significant charge sectors, each contributing $\sim \sqrt{\log(t)}$. We get satisfying results comparing the prediction above to numerical results.
	As for the charge imbalance resolved negativity, the CFT results involve hard-to-calculate conformal blocks, but we can still derive approximate expressions which qualitatively match the numerical results.
	
	The rest of the paper is organized as follows: In Sec. \ref{sec:CFT} we present the theoretical background for the calculation of entanglement entropies and negativities for 1+1D CFT, and extend this method for calculating the charge resolved entanglement entropies and charge imbalance resolved negativities. We then use this method to calculate the charge resolved vNEEs and imbalance resolved RNs after a local quench. In Sec. \ref{sec:results} we compare the CFT predictions to numerical results for the XX and XXZ model. We summarize our findings and outline future directions in Sec. \ref{sec:conclusion}. Appendices \ref{app:entropy} and \ref{app:neg} include the full formulas for the symmetry resolved entanglement entropies and negativities, as these are too cumbersome to appear in full in the main text.
	
	\section{\label{sec:CFT}Conformal Field Theory Analysis}
	\subsection{\label{sec:cftVNEE} The Entanglement Entropy}

	We will first recap the calculation of the total REs of a 1+1D CFT in the ground state and after a local quench~\cite{CC}, and then show how to generalize these techniques to the charge-resolved REs. The calculation is based on the replica trick. Space-imaginary time will be represented by the complex plane, with $z=x+iv \tau$, with $v$ the velocity of excitations. Representing the DM as a path integral, the RDM (when the total system is in its ground state) is represented as a path integral over the complex plane with a cut at $\tau=0$ at subsystem $A$'s coordinates. Different boundary conditions on this cut give different matrix elements of the RDM, see Fig \ref{fig:rdm} for the geometric representation. Sewing $n$ copies of the RDM together is equivalent to the multiplication of the n copies of the RDM, and by doing so one obtains the $n$th RE as a path integral on an $n$-sheet Riemann surface, see Fig. \ref{fig:sub_replica}. The transition between the copies is effected by \textit{twist field operators} $\mathcal{T}_n$ and $\tilde{\mathcal{T}}_n$. $\mathcal{T}_n$ and $\tilde{\mathcal{T}}_n$ transfer particles from one copy to the next clockwise and counterclockwise, respectively.  Hence, the RE is proportional to the correlation function of these twist fields:
	\begin{equation}
	S_A^{(n)} \propto \left\langle \mathcal{T}_n(x = x_{L}, \tau = 0) \tilde{\mathcal{T}}_n(x = x_{R}, \tau = 0) \right\rangle,
	\end{equation}
	for $x = x_{L}$ and $x = x_{A}$ the endpoints of subsystem $A$ (taken as a single interval). The scaling dimension of the twist fields was calculated in Ref.~\onlinecite{CC}:
	\begin{equation}
	d_n = \bar{d_n} = \frac{c}{24}\left(n - n^{-1}\right),
	\end{equation}
	and the resulting RE and vNEE, again from Ref.~\onlinecite{CC}, are:
	\begin{equation}\label{eq:ee}
	\begin{aligned}
	& S_A^{\left(n\right)} = \Tr \rho_A^n = c_n \left(\frac{L_A}{a}\right)^{-c\left(n - n^{-1}\right)/6}, \\
	& S_A = \frac{c}{3}\log\frac{L_A}{a} + c_1^\prime,
	\end{aligned}
	\end{equation}
	where $L_A = |x_{L} - x_{R}|$ is the length of subsystem $A$, $a$ is a cutoff corresponding to, e.g., a lattice spacing, $c$ is the conformal central charge, and $c_n$ and $c_1^\prime$ are constants that cannot be predicted using CFT.

\begin{figure*}[ht]
	\subfloat
	[\label{fig:tRDM}]{		\includegraphics[width=.3\linewidth]{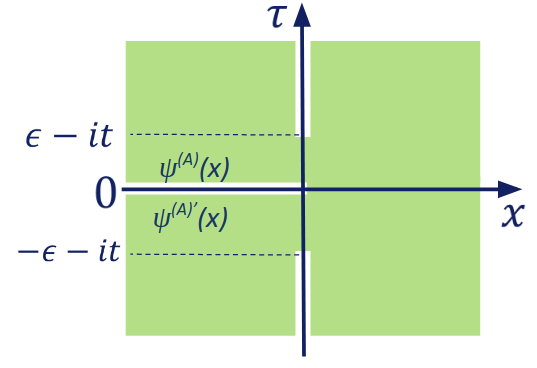}%
	}$\rightarrow$ \hfill
	\subfloat
	[\label{fig:tTransform}]{		\includegraphics[width=.22\linewidth]{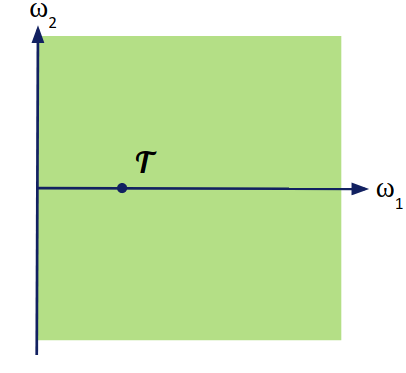} $\rightarrow$
	}
	\subfloat
	[\label{fig:bcft}]{		\includegraphics[width=.38\linewidth]{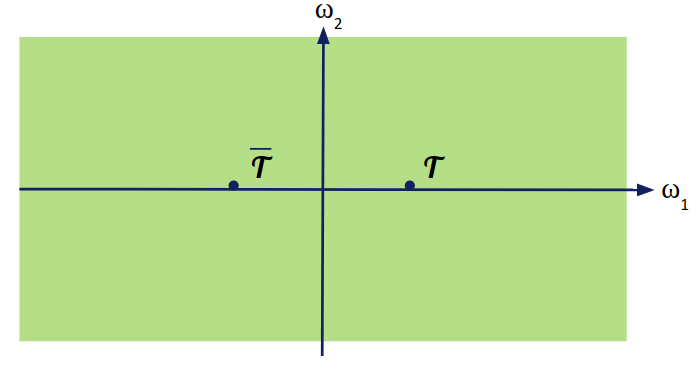}%
	}
	\caption{(a) A geometric representation of the time dependent RDM element after a local quantum quench. A  twist field operator can be added to simulate the replica trick. (b) We map the cut plane from (a) to the right half plane using the conformal transformation (\ref{eq:transCC}). (c) ``Unfolding'' of the system from the right half plane to the full plane, doubling the number of operators.}
	\label{fig:replicaT}
\end{figure*}

	Let us now go on to the time-dependent local quench scenario. We prepare two identical systems in the ground state, and at time $t = 0$ we couple them at one contact point. For this time dependent case, the geometry representing the RDM is slightly more complicated: The two halves of the system are not connected before $t=0$. Upon analytical continuation to imaginary time, this is expressed as slits in each sheet, parallel to the imaginary time axis. The slits are separated by a small gap $\epsilon$, which serves as a convergence factor. This is shown in Fig. \ref{fig:tRDM}. We then use the conformal transformation~\cite{CC}:
	\begin{equation}\label{eq:transCC}
	\omega(z) = \frac{z}{\epsilon} + \sqrt{\left(\frac{z}{\epsilon} \right)^2 +1},
	\end{equation}
	which takes the Riemann sheet with the slits into the right half plane, as demonstrated in Fig. \ref{fig:tTransform}.

	A system living on the right half plane has a boundary on the imaginary time axis, and requires the use of boundary CFT (BCFT)~\cite{BCFT}: We separate our field into a holomorphic part and an anti-holomorphic part, and ``unfold'' the anti-holomorphic part into the left half plane. This also duplicates the operators in our system, turning an $n$-point function into a $2n$-point function, as demonstrated in Fig. \ref{fig:bcft}. For an infinite system with one boundary point between the systems $A$ and $B$ we are then left with a two-point function for the composite twist fields. The final result for the entanglement when $A$ and $B$ are the two parts of the system brought together by the local quench, is:

	\begin{equation}\label{eq:tee}
	\begin{aligned}
	& S_A^{\left(n\right)}\left(t\right) \propto \left(\frac{\epsilon^2 + (vt)^2}{a\epsilon/2}\right)^{-c\left(n - n^{-1}\right)/12}, \\
	& S_A(t) = \frac{c}{3}\log\frac{vt}{a} + 	k_0,
	\end{aligned}
	\end{equation}
	where the last equation applies for $vt \gg \epsilon$, and $k_0$ is a non-universal constant. We can define the effective length $L_\mathrm{eff}(t) = \sqrt{\epsilon^2 + (vt)^2}$ and notice that the REs take the form $S^{(n)}_A = \tilde{c}_n\left(\frac{L_\mathrm{eff}(t)}{a}\right)^{d_n}$, which is equivalent to Eq. (\ref{eq:ee}) with $L_A \rightarrow L_\mathrm{eff}(t)$. One can think of the entanglement as carried by quasiparticles with velocity $v$, propagating from the quench point to the rest of the system. $L_\mathrm{eff}(t)$ represents the parts of the system reached by these quasiparticles. Even for more complicated geometries and conformal transformation, the result will always be of the form above, with $L_\mathrm{eff}(x, t)$ modified according to the geometry.
	
	For finite subsystems $L_A =  L_B = L/2$, one may use the following conformal transformation~\cite{SD}:
	\begin{equation}\label{eq:transSD}
	\begin{aligned}
	& \omega(z) = \coth\left(\frac{\pi\epsilon}{2L}\right)\frac{1+\zeta(z)}{1-\zeta(z)}, \\ & \quad \text{with}\quad \zeta(z) = \sqrt{\frac{\sinh\frac{\pi}{L}(z + \epsilon)}{\sinh\frac{\pi}{L}(z - \epsilon)}}.
	\end{aligned}
	\end{equation}
	The effective lengths in this case, with the quench point at the boundary point between the subsystems or shifted from it, appear in Appendix \ref{app:entropy}.
	
	Let us now proceed to the calculation of the charge-resolved entropy in the ground state. In Ref.~\onlinecite{GS}, the twist fields are multiplied by a vertex operator $\mathcal{V}(\alpha)$. For a charge $\hat{N}$ related to an abelian $U(1)$ symmetry, the vertex operator is coupled to the charge such that a particle going counter-clockwise around $\mathcal{V}(\alpha)$ will acquire a phase $e^{i\alpha}$. If the vertex operators are placed at the edges of subsystem $A$, as the twist fields are, the resulting correlation function will give:
	\begin{equation}\label{eq:fluxResolved}
	S_A^{(n)}(\alpha) = \Tr \rho_A^n e^{i\hat{N}_A \alpha}.
	\end{equation}
	We can think of this vertex operator as introducing a flux $\alpha$ for the particles, and define the measure in (\ref{eq:fluxResolved}) to be the \textit{flux resolved RE}, see Fig. \ref{fig:replicaFlux}.
	A Fourier transform will lead us to the charge resolved RE,
	\begin{equation}\label{eq:fourier}
	S_A^{(n)}(N_A) = \int_{-\pi}^{\pi}\frac{d\alpha}{2\pi} e^{-iN_A \alpha} S_A^{(n)}(\alpha).
	\end{equation}
	
	For concreteness, let us consider the case of gapless 1D spin-$\frac{1}{2}$ chain, or, equivalently (via the Jordan-Wigner transformation~\cite{JW}), gapless 1D interacting fermion chain. According to the Luttinger liquid paradigm~\cite{vD, Senechal}, the system can be described both in terms of fermions and in terms of noninteracting bosons. In the spinless case, the relation between the fermion field $\psi$ and the boson field $\phi$ is $\psi \propto e^{i\phi}$. The bosonic field is governed by the gapless non-interacting Hamiltonian density:
		\begin{equation}
		\begin{split}
		& H = \frac{1}{2}v\left(K (\Pi)^2 + \frac{1}{K}(\partial_x \phi)^2\right),
		\end{split}
		\end{equation}
		where $\Pi$ is the canonically conjugate momentum to $\phi$, and $v$ and $K$ are model-dependent parameters. $v$ is the field velocity. $K$ is called the \textit{Luttinger parameter},  $K=1, K < 1$ and $K>1$ corresponding to noninteracting, repulsive and attractive fermions, respectively.

\begin{figure*}[t]
	\subfloat{
		\includegraphics[width=0.3\linewidth]{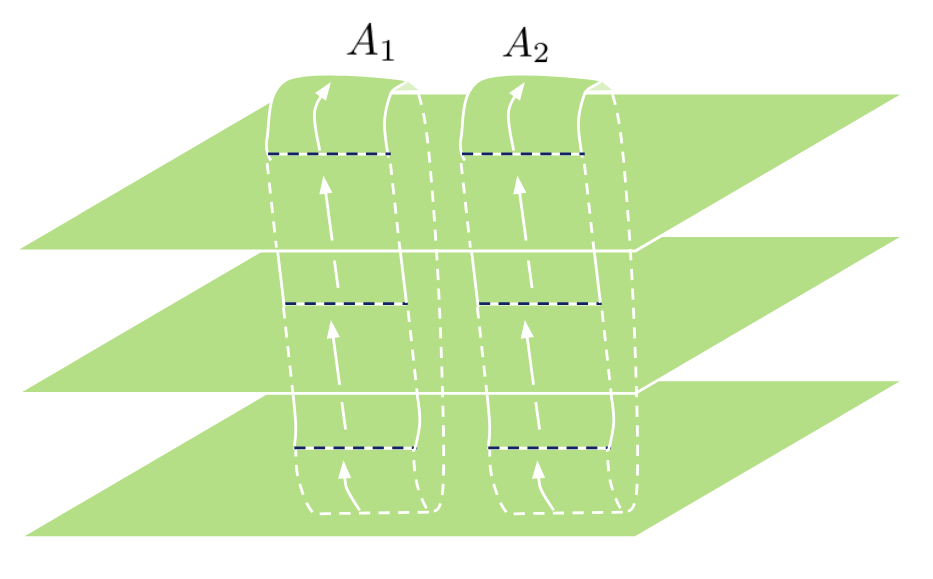}	
	}
	\subfloat{
		\centering
		\includegraphics[width=0.3\linewidth]{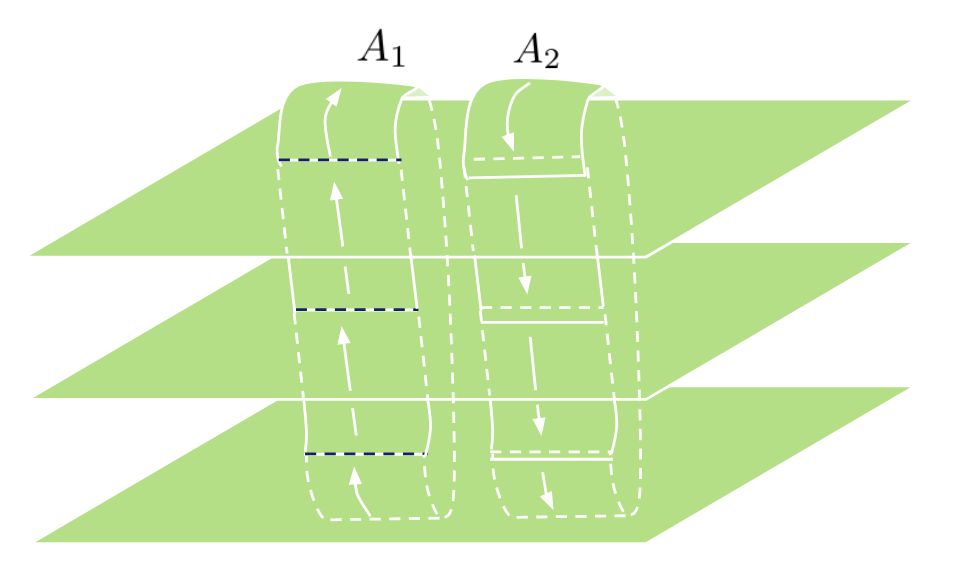}	
	}
	\subfloat{
		\centering
		\includegraphics[width=0.33\linewidth]{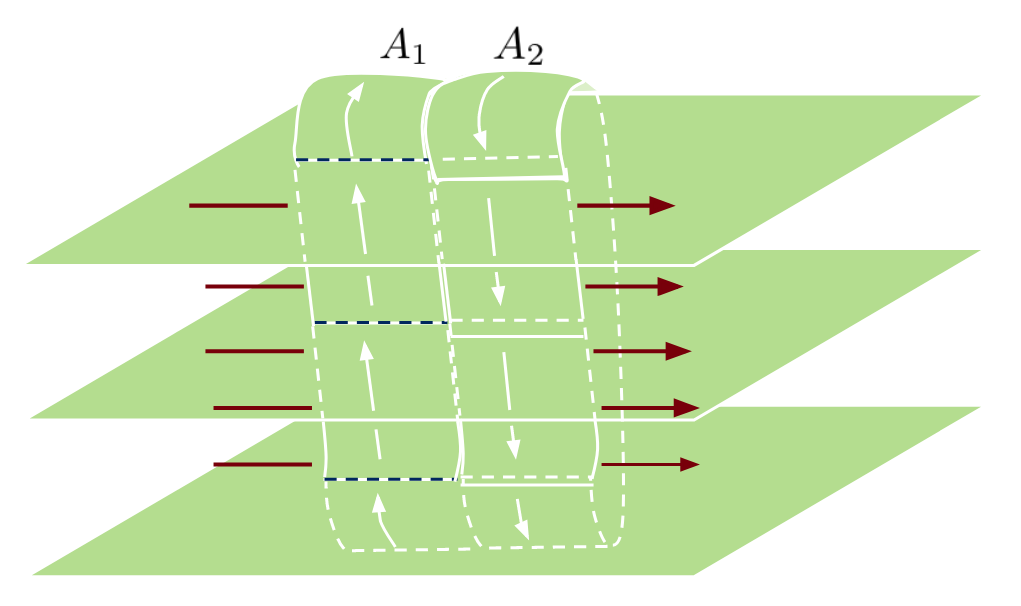}
	}
	\caption{(a) Geometric representation of $\Tr\left(\rho_{A_1 \cup A_2}\right)^n$, equivalent to figure  \ref{fig:replica}. The subsystems are disjoint for clarity of the figure, but in this work we only examined the case of adjoint subsystems. (b) Geometric representation to $\Tr\left(\rho_{A_1 \cup A_2}^{T_2}\right)^n$. The partial transposition is portrayed by reversing the direction of the sewing between copies in $A_2$. (c) A flux $\alpha$ is added to the partially transposd replica trick. Particles in subsystem $A_1$ will go around the flux clockwise, and acquire a phase $\alpha$. Particles in subsystem $A_2$ will go around the flux counter-clockwise and acquire a phase $-\alpha$, resulting in an overall phase $\hat{Q}\alpha$.
		\label{fig:replicaNeg}}
\end{figure*}

\begin{figure*}[t]
	%		\subfloat
	%		[\label{fig:exactEE}]{		\includegraphics[width=.639\linewidth]{resExactEEBoth.png}}
	\subfloat
	[\label{fig:exactEE}]{		\includegraphics[width=.5\linewidth]{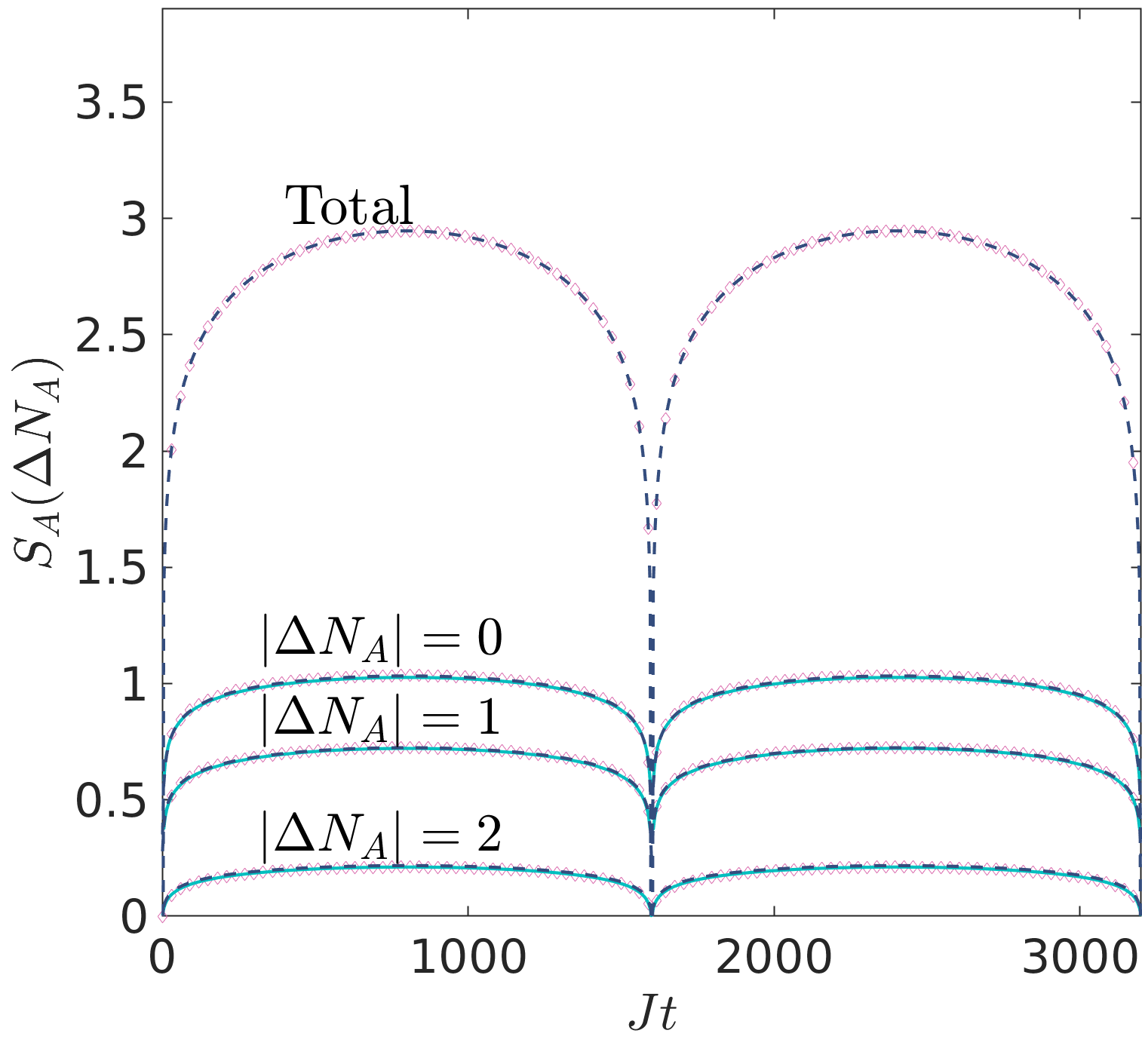}}
	\subfloat
	[\label{fig:exactEEShift}]{		\includegraphics[width=.5\linewidth]{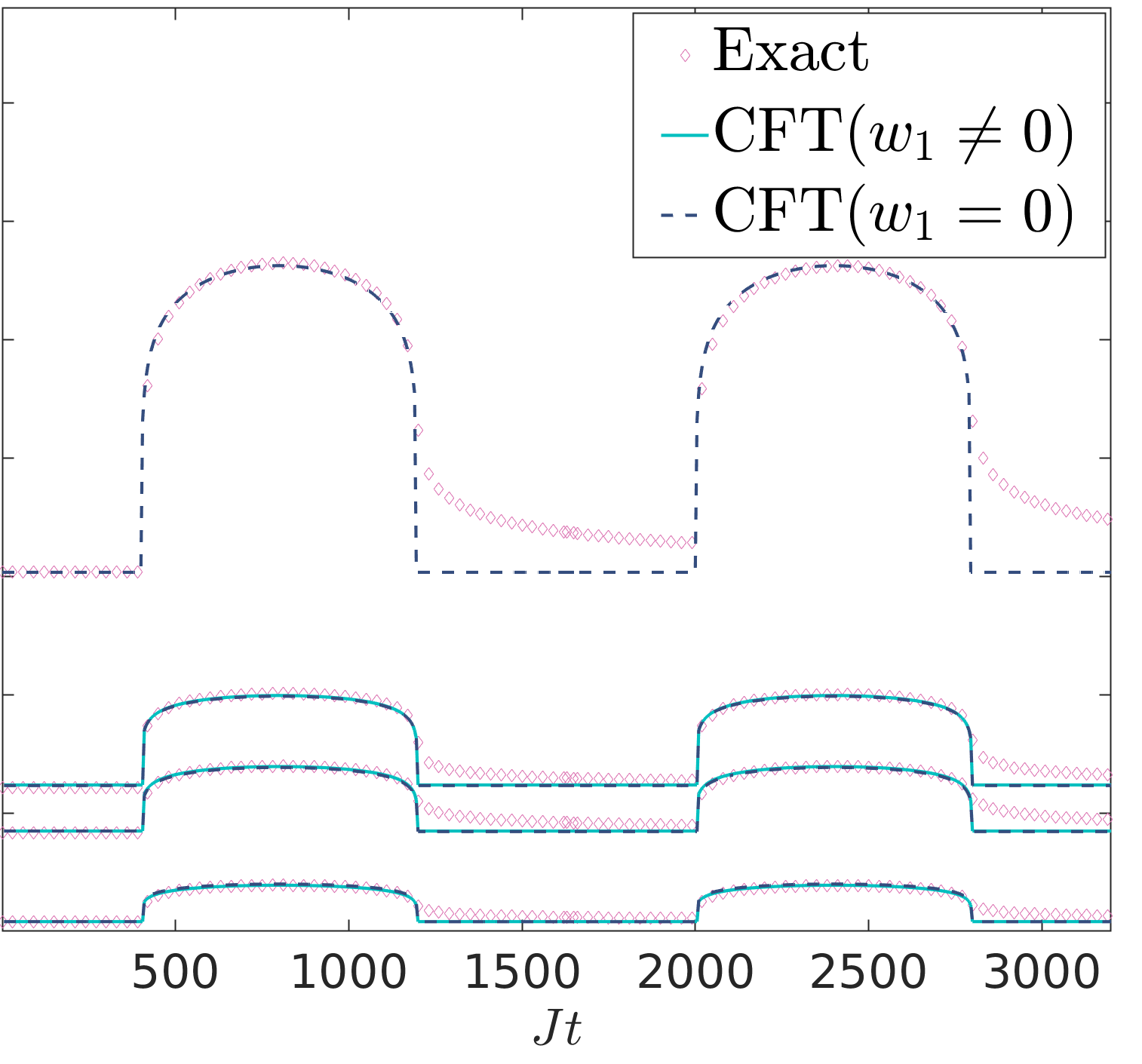} 
	}
	\\
	\subfloat
	[\label{fig:exactEEAlpha}]{		\includegraphics[width=.5\linewidth]{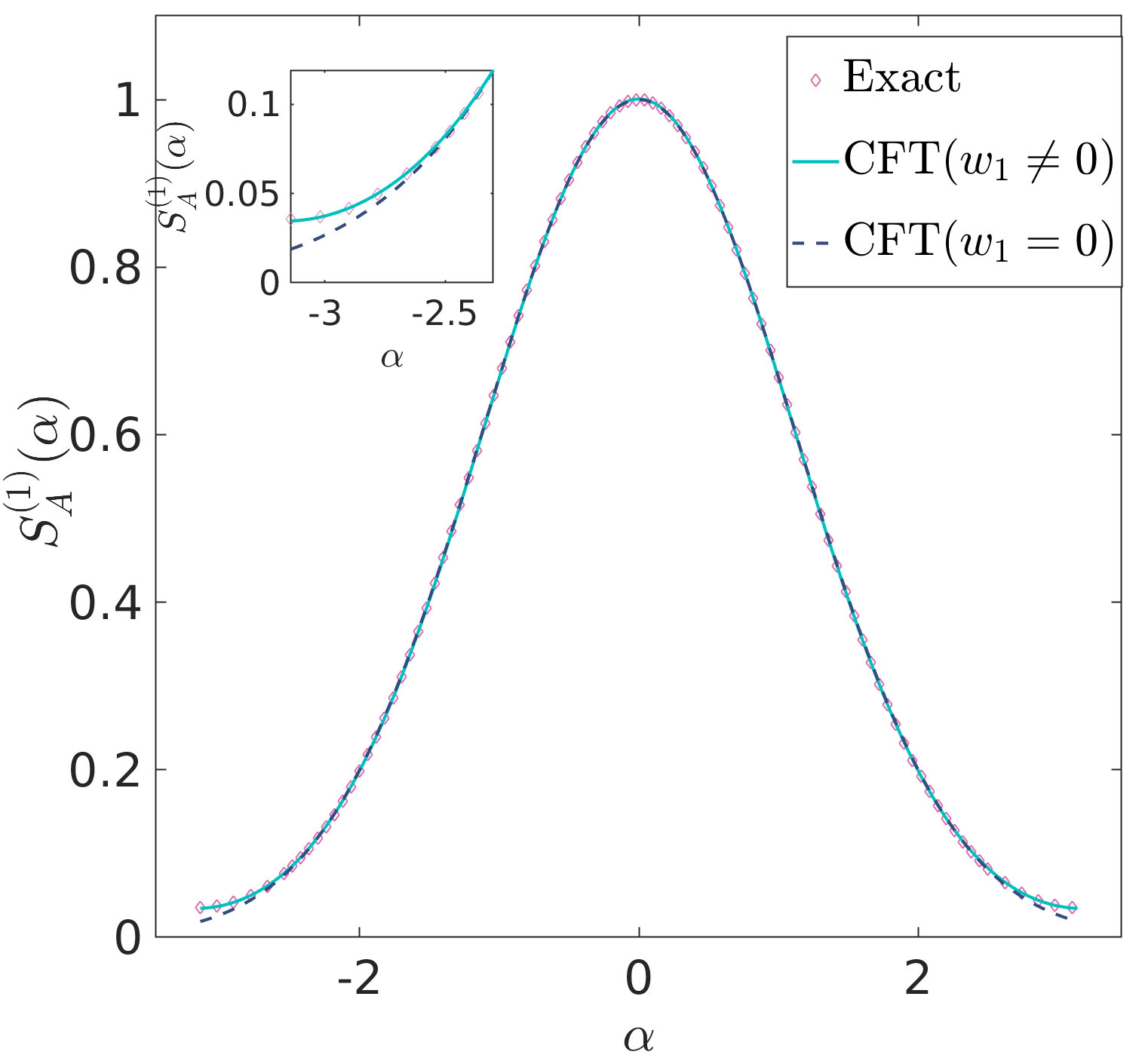}%
	}
	\caption{Exact results for the charge resolved entanglement of the XX model with site number $L/a = 3200$ sites. In (a) the boundary between the subsystems $A$ and $B$ is the quench point, and in (b) the boundary is shifted by $L/4a$ from the quench point. In (c) we present the fit of the flux resolved 1st RE for some generic $Jt = 195$. Inset: Zoom-in on the region in which the difference between the results for $w_1=0$ and $w_1 \ne 0$ (defined in Eq. (\ref{eq:wpm})) is noticeable for $\alpha$ close to $-\pi$.}
	\label{fig:resExact}
\end{figure*}
	
	The boson-fermion relation above allows us to define the vertex operator. We annotate the bosonic field
		living on the the $j$th plain (i.e., $j$th copy of the system)
		by $\phi_j$. The vertex operator can be chosen as
	\begin{equation}\label{eq:vertexOpp}
	\mathcal{V} = e^{i\frac{\alpha}{2\pi}\phi_j},
	\end{equation}
	with a scaling dimension derived in Ref.~\onlinecite{GS}:
	\begin{equation}\label{eq:vertexScaling}
	d^{(\mathcal{V})} = \frac{K}{2}\left(\frac{\alpha^2}{2\pi}\right).
	\end{equation}

		\begin{figure*}[t]
		\centering
		\subfloat[\label{fig:MPS}]{
			\includegraphics[width=0.26\linewidth]{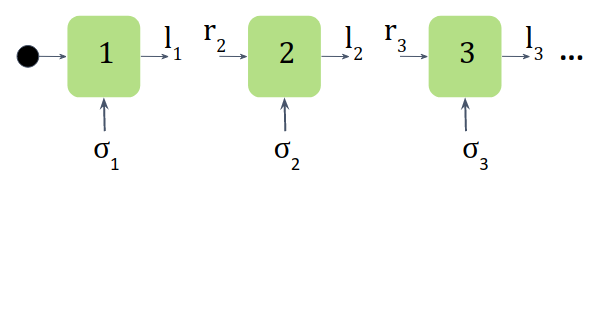}	
		} 				
		\subfloat[\label{fig:dmrgneg1}]{
			\includegraphics[width=0.19\linewidth]{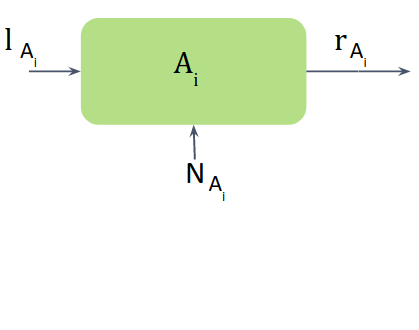}	
		} 		
		\subfloat[\label{fig:dmrgneg2}]{
			\includegraphics[width=0.26\linewidth]{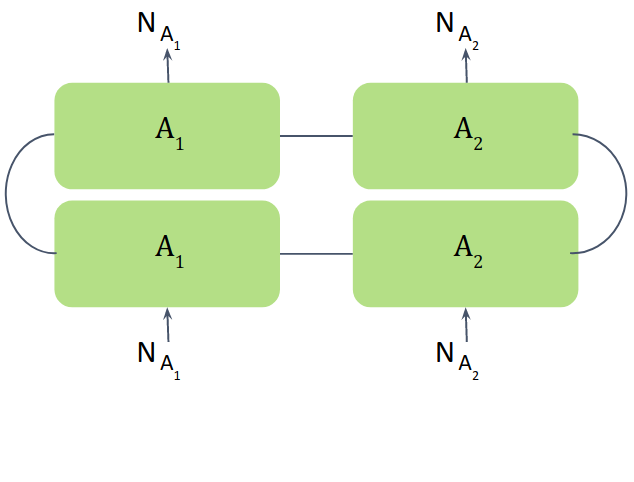}	
		}
		\subfloat[\label{fig:dmrgneg3}]{
			\includegraphics[width=0.285\linewidth]{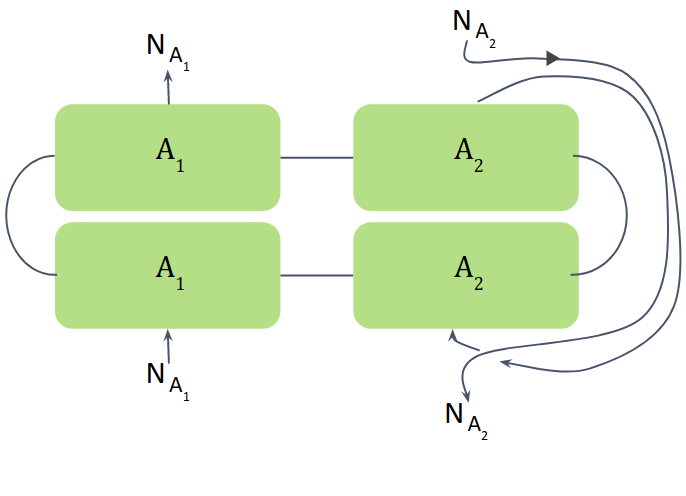}	
		}
		\caption{(a) (Based on Ref.~\onlinecite{SCHOLLWOCK}) In the MPS representation, we represent the state as rank-3 tnsors, one for each site in our system. Here we have a representation of the leftmost 3 sites. The bottom 'legs' represent the physical index of the site, and the left and right 'legs' represent the connection of the sites to the rest of the system from left and right. Adjacent indexes ($l_i$ and $r_{i+1}$) are contracted together. On the left of the leftmost site we add a dummy tensor representing a Hilbert space of size 1. We can separate our tensors into blocks corresponding to different charge sectors. The arrows on the legs represent ingoing or outgoing charge to the tensor such that the total charge sums up to zero for each block. For example, in the second tensor, the charge on the chain of the right to site 2, index $r_2$, and the charge on the physical index, $\sigma_2$, sum up to the outgoing charge to the left of the chain, $l_2$. (b) A tensor representing a subsystem, obtained by contracting all its single-site tensors. The bottom leg corresponds to the combined physical index. The charge corresponding to the physical index is the charge on the entire subsystem, $N_{A_i}$. (c) By taking two copies of the tensor for each system, create the RDM by contracting the non-physical legs tracing out the environment. (d) Applying an explicit transpose on the legs corresponding to subsystem $A_2$ in order to get the partially transposed RDM. We can see that by performing the partial transposition, the charge on the physical index of subsystem $A_2$ is switched from ingoing to outgoing on the bottom indices, and vice versa on the top indices. This results in matrix blocks corresponding to the charge imbalance rather than the charge. (panels (b-d) are based on Ref.~\onlinecite{RAC}).}
		\label{fig:dmrgNeg}
	\end{figure*}

	One may now derive an expression for the scaling function of the composite twist field ~\cite{GS}:
	\begin{equation}\label{eq:scalingD}
	d_n(\alpha) = \frac{c(n -n^{-1})}{24} + \frac{d^{(\mathcal{V})}}{n}.
	\end{equation}
	This leads to Gaussian dependence of the charge-resolved RE on the charge, provided $\ln (L_A/a) \gg 1$:
	\begin{equation}\label{eq:gsRes}
	\begin{aligned}
	S_A^{(n)}(N_A) \cong S_A^{(n)}(\alpha = 0) \sqrt{\frac{\pi n}{2 K \ln L_A/a}}e^{-\frac{\pi^2 n \Delta N_A^2}{2K\ln L_A/a}},
	\end{aligned}
	\end{equation}
	where $\Delta N_A = N_A - \left\langle \hat{N}_A \right\rangle$, $\left\langle \hat{N}_A \right\rangle$ being the expectation value of $\hat{N}_A$ in the ground state. For different geometries, one can replace $L_A$ in Eq. (\ref{eq:gsRes}) by the suitable effective length $L_\mathrm{eff}$. The Gaussian dependence on the charge is a result of the unnormalized form of the RDM blocks. In Refs~\onlinecite{EquipS, EquipC}, the blocks are normalized and a the different normalized blocks rise to equal entropies, a property these works refer to as ``equipartition''.

	These results could be extended to smaller subsystems. We notice that the most general form for $\mathcal{V}$ is in fact:
	\begin{equation} \label{eq:wpm}
	\mathcal{V} = \sum_{m = -\infty}^{\infty} w_{|m|} e^{i\left(\frac{\alpha + 2\pi m}{2\pi}\right)^2 \phi},
	\end{equation}	
	where $m$ is integer and $w_{|m|}$ is the corresponding weight.
	For a large enough $L_\mathrm{eff}$, the zeroth order term ($w_{|m|} = 0$ for $m \ne 0$) is sufficient, since its correlation function features the slowest decay with subsystem size. In the time dependent case (to be discussed shortly), we often find it necessary to include the next order, $w_1\ne 0$, since its contribution is becoming important for short times. For the XX model studied below, the parameter $w_1$ can be extracted for the ground state using similar methods to those employed in Ref.~\onlinecite{CalEs}. However, in the time dependent case no such results are available, and we resort to extracting $w_1$ from a fit to our numerical results.

	Having laid out all the necessary groundwork, we may now derive our new results for the charge resolved entropies following a local quench. Combining Eqs. (\ref{eq:tee}), (\ref{eq:vertexScaling}) and (\ref{eq:wpm}), we find the following expression for the dynamics of the flux resolved RE after a local quench:
	\begin{equation}\label{eq:tEEFlux}
	\begin{aligned}
	& S^{(n)}_A(\alpha, t) = S^{(n)}_A(\alpha = 0, t) \times \\
	& \sum_{m = -\infty}^\infty w_m \left(\frac{L_\mathrm{eff}}{a}\right)^{-\frac{c}{12}(n - n^{-1}) - \frac{K}{n}\left(\frac{\alpha + 2\pi m}{2\pi}\right)^2}.
	\end{aligned}
	\end{equation}
	Plugging this into Eq. (\ref{eq:fourier}) will give us the charge resolved RE. For a finite system as studied below, we use the transformation (\ref{eq:transSD}) and get the relevant expression $L_\mathrm{eff}$ to be substituted in (\ref{eq:tEEFlux}). For the case where the boundary between subsystems $A$ and $B$ is shifted from the quench point, we use the same conformal transformation (\ref{eq:transSD}), but place the twist fields away from the slits~\cite{SD}, and obtain a modified expression for $L_\mathrm{eff}(t)$. The final formulas we used appear in Appendix \ref{app:entropy}.
	%%%

	\subsection{The Entanglement Negativity}\label{sec:cftNeg}

	The replica trick for the negativity is derived in Ref.~\onlinecite{CCT}, and a geometric representation for it appears in Fig. \ref{fig:replicaNeg}. For two adjacent single-interval subsystems $A_1 = [x_{L}, x_{B}], A_2 = [x_{B}, x_{R}]$, twist field operators are added at the three boundary points $x_{L}, x_{B}, x_{R}$ such that:
	\begin{equation}\label{eq:renyiNegTotal}
	\begin{aligned}
	\mathcal{N}_{A_1,A_2}^{(n)} \propto \left\langle \mathcal{T}_n(x_{L}, 0) \tilde{\mathcal{T}}^2_n(x_{B}, 0) \mathcal{T}_n(x_{R}, 0)\right\rangle.
	\end{aligned}
	\end{equation}
	
	Calculating the negativity for the local quench case requires introducing these twist fields into $n$-copies of the geometry depicted in Fig. \ref{fig:replicaT}. We are again forced to use BCFT and double the number of operators. We are left with calculating a 6-point function. For a $j$-point function with $j \ge 4$, the result can be predicted by CFT only up to a nonuniversal function $\mathcal{F}$,~\cite{BYB} which depends on the full operators content of the studied theory. We thus restrict ourselves to limits in which $\mathcal{F}$ is approximately constant. It is for this reason that we do not study the case where $A_1$ and $A_2$ are disjoint --- following the arguments above, we will need to calculate an 8-point function, for which the effect of nonuniversal function $\mathcal{F}$ is expected to be even more significant.

	In Ref.~\onlinecite{WCR} the RN after a local quench was found to be:
	\begin{equation}\label{eq:renyiNegT}
	\mathcal{N}_{A_1,A_2}^{(n)} = \prod_i \left|\frac{d\omega}{dz}\right|^{d_i}_{z_i} \langle{\mathcal{T}_n(\omega_L)\tilde{\mathcal{T}}_n^2(\omega_B)\mathcal{T}_n(\omega_R)}\rangle,
	\end{equation}
	where $\omega_i = \omega(x_i)$ for $i = L,B,R$, as defined in Eq. (\ref{eq:transCC}) or (\ref{eq:transSD}). The unfolded 6-point function was found to be:
	\begin{equation} \label{eq:nuF}
	\begin{aligned}
	&\langle{\mathcal{T}_n(\omega_1)\tilde{\mathcal{T}}_n^2(\omega_2)\mathcal{T}_n(\omega_3)}\rangle = \\ & \quad \prod_i \frac{1}{|\omega_i - \tilde{\omega_i}|^{d_{n, i}}} \left(\frac{\eta_{1,3}^{d_n^{\left(2\right)} - 2d_n}}{\eta_{1,2}^{d_n^{\left(2\right)}} \eta_{2,3}^{d_n^{\left(2\right)}}}\right)^{1/2}\mathcal{F}\left(\{\eta_{j,k} \}\right),
	\end{aligned}
	\end{equation}
	where $d_n^{(2)}$ is the scaling dimension of $\mathcal{T}_n^2$ calculated in Ref.~\onlinecite{CCT}, and is different for $n_e$ and $n_o$, even and odd values of $n$, respectively: $d_n^{(2)} = \frac{c}{12}\left(\frac{n_e}{2} - \frac{2}{n_e}\right)\quad d_n^{(2)} = \frac{c}{24}\left(n_o - \frac{1}{n_o}\right)$.  In addition,  $\tilde{\omega}_i = -\omega_i^*$ and $\eta_{i,j} = \frac{\left(\omega_i - \omega_j\right)\left(\tilde{\omega}_i - \tilde{\omega}_j\right)}{\left(\tilde{\omega}_i - \omega_j\right)\left(\omega_i - \tilde{\omega}_j\right)}$. $\mathcal{F}$ approaches a constant value for $vt \ll l, vt = l + 0^+$, and $vt > l$ for two identical subsystems $L_{A_1}=L_{A_2}=l$, which is the case on which we focus here. We will only consider $l \ll L$, and so neglect corrections due to boundary conditions. 
	
	The charge-imbalance resolved RNs are again obtained from the flux resolved RNs,
	\begin{equation}\label{eq:rnAlpha}
	\mathcal{N}_{A_1,A_2}^{(n)}(\alpha) = \Tr((\rho_A^{T_2})^n e^{i\hat{Q}\alpha}),
	\end{equation}
	\begin{equation}
	\mathcal{N}_{A_1,A_2}^{(n)}(Q) = \int_{-\pi}^{\pi} \frac{d\alpha}{2\pi} \mathcal{N}_{A_1,A_2}^{(n)}(\alpha) e^{-i\alpha Q}.
	\end{equation}
	We obtain the flux resolved RNs by adding vertex operators at the boundaries between the subsystems\cite{CGS}. The additivity of the scaling dimensions, Eq. (\ref{eq:scalingD}), results in
	\begin{equation}\label{eq:renyiNegCGS}
	\frac{\mathcal{N}_{A_1,A_2}^{(n)}(\alpha)}{\mathcal{N}_{A_1,A_2}^{(n)}} \propto \left\langle \mathcal{V}(\alpha,x_{L}, 0)  \mathcal{V}(-2\alpha, x_{B}, 0) \mathcal{V}(\alpha, x_{R}, 0)\right\rangle.
	\end{equation}

	We can now present our new results. Combining Eqs. (\ref{eq:transCC}) and (\ref{eq:renyiNegCGS}), and using the vertex operators correlation function from Ref.~\onlinecite{BYB}, we obtain the flux resolved RN for two adjacent systems with an infinite environment (we present the expression for the simple case, $w_{|m|>0}=0$)
	\begin{equation}\label{eq:tRNFlux}
	\begin{aligned}
	& \frac{\mathcal{N}_{A_1,A_2}^{(n)}(\alpha, t)}{\mathcal{N}_{A_1,A_2}^{(n)}(t)} = \prod_{i, j} e^{-\frac{\alpha_i\alpha_j}{(2\pi)^2}\ln\left(\left|\omega^\prime_i - \omega^\prime_j\right|\right)},
	\end{aligned}
	\end{equation}
	where $i$ and $j$ count the 6 positions of the vertex operators, $\omega^\prime_i  = \omega_L, \omega_B, \omega_R, \tilde{\omega}_L, \tilde{\omega}_B, \tilde{\omega}_R$ and $\alpha_i= \alpha(x_i) = \alpha, -2\alpha, \alpha, -\alpha, 2\alpha, -\alpha$, respectively. To the first order (and taking $w_m = 0$ for $|m| > 1$ in Eq. (\ref{eq:wpm})), we get
	\begin{equation}\label{eq:tRNFluxWpm}
	\begin{aligned}
	& \frac{\mathcal{N}_{A_1,A_2}^{(n)}(\alpha, t)}{\mathcal{N}_{A_1,A_2}^{(n)}(t)} = \sum_{\mu} \prod_{i, j} e^{-\frac{\alpha_i^{(\mu)}\alpha_j^{(\mu)}}{(2\pi)^2}\ln\left(\left|\omega^\prime_i - \omega^\prime_j\right|\right)},
	\end{aligned}
	\end{equation}
	where $\mu = 1 \cdots 6$ , and $\alpha_i^{(\mu)} = \alpha_i + 2\pi\delta_{i, \mu}  - 2\pi\delta_{i+3, \mu} $. The explicit expressions for $\omega^\prime_i - \omega^\prime_j$ were derived in Ref.~\onlinecite{WCR}, and are reproduced for completeness in Appendix \ref{app:neg}.

	\begin{figure*}[t]
		\centering
		\includegraphics[width=.98\linewidth]{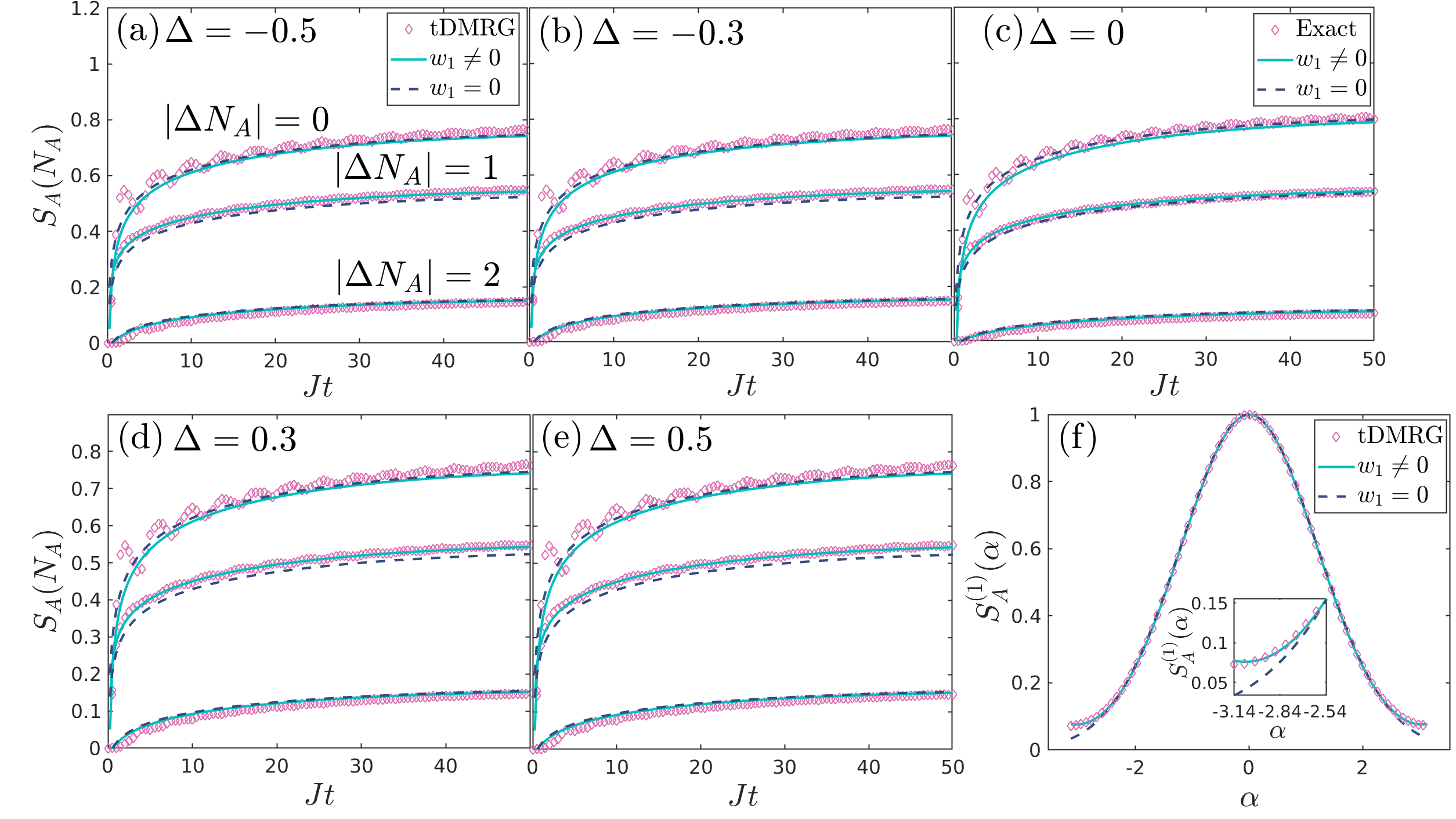}%
		\caption{(a, b, d, e) tDMRG results for the charge-resolved entanglement of the XXZ model with different values of $\Delta$, for a system with $L/a = 256$ sites. For comparison, exact diagonalization results for a system of the same size in the XX model are presented in (c). In (f) we present the fit of the flux resolved 1st RE for a generic time $Jt=26$. Inset: Zoom-in on the region near $\alpha = -\pi$.}
		\label{fig:tdmrgEE}
	\end{figure*}
	%%%
	%\begin{figure}[t]
	%	\centering
	%	\includegraphics[width=.98\linewidth]{vNEE_NI.png}%
	%	\caption{tDMRG results for a NNN interaction model (Eq. (\ref{eq:ni})) with $\Delta = -0.5, \Delta^\prime_2 = 0.02$, and $\Delta_2 = 0.01$ and a system with $L/a = 64$ sites. The parameters were chosen such the system remains in the gapless phase~\cite{JL}.}
	%	\label{fig:ni}
	%\end{figure}
	%%%

	\section{\label{sec:results}Numerical Results}
	\begin{figure*}[t]
		\centering
		\includegraphics[width=1\linewidth]{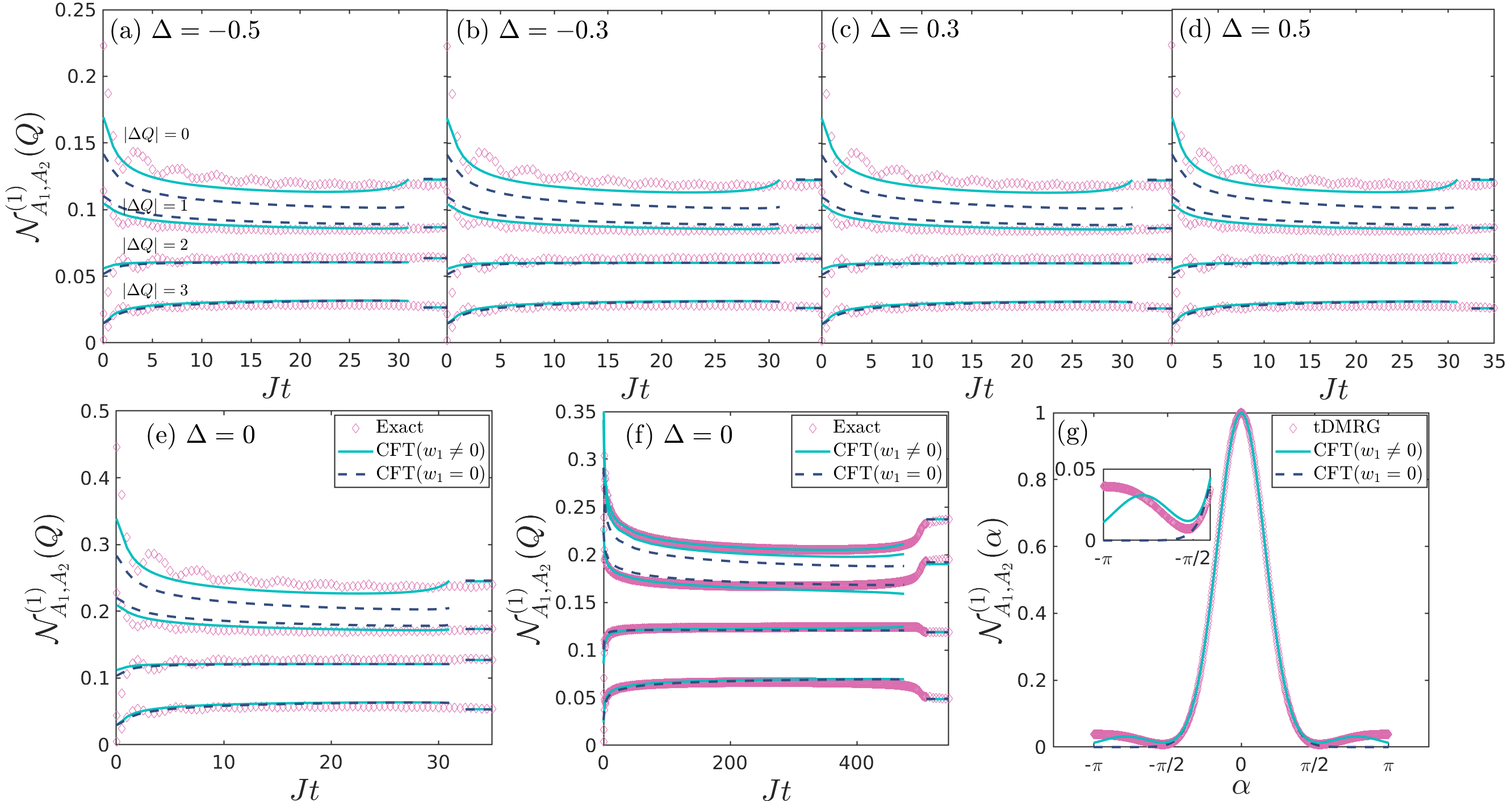}
		\caption{The first imbalance resolved RN. (a-d) tDMRG results for the XXZ model with different values of $\Delta$ for $\frac{L}{a} = 256$ and $\frac{L_{A_i}}{a} = 64$. (e) Exact results for a system with the same size as the tDMRG results for comparison. (f) The exactly solvable XX model for a very large system ($\frac{L}{a} = 10,000$), for $\frac{L_{A_i}}{a} = 1000$.  For $t \approx \frac{L_{A_i}}{v}$ the conformal approximation is not applicable, and these region were omitted. (g) A fit of the flux resolved first RN for for some generic $Jt=6$ for the XXZ model with $\Delta = -0.5$. Inset: Zoom-in on the region near $\alpha = -\pi$.}
		\label{fig:negR1}
	\end{figure*}
	\subsection{\label{sec:resVNEE} Entanglement Entropy}
	
	\textit{The XXZ model}. 
	We compare our CFT predictions to numerically obtained results for the XXZ spin chain,
	\begin{equation}\label{eq:XXZ}
	H = J\sum_{n}\left(\sigma^+_n \sigma^-_{n+1} + \text{h.c.} + \Delta \sigma^z_n \sigma^z_{n+1}\right),
	\end{equation}
	when $\sigma^+, \sigma^-, \sigma^z$ are the Pauli matrices. Using the Jordan Wigner tranformation~\cite{JW}
	\begin{equation}\label{eq:jordanWigner}
	\begin{split}
	& \sigma_i^{+} = f_i^\dagger, \\
	& \sigma_i^{-} = f_i, \\
	& \sigma_i^{z} = 2f_i^\dagger f_i - 1, \\
	& c_i = e^{-i\pi\sum_{j < i} f_j^\dagger f_j} f_i,
	\end{split}
	\end{equation}
	the XXZ model can be interpreted as a spinless fermionic chain.
	\begin{equation}\label{eq:XXZf}
	\begin{split}
	H = & J\sum_{n} \bigg( c^\dagger_n c_{n+1} + \text{h.c.} +   \\ 
	&  \Delta (2 c^\dagger_n c_n - 1)(2c^\dagger_{n+1} c_{n+1} - 1)\bigg),
	\end{split}
	\end{equation}
	where the $c_n$ annihilation operators obey the fermionic anti-commutation relations. The system is a gapless Luttinger liquid for $-1 < \Delta \le 1$. Its Luttinger parameter and velocity can be extracted from the Bethe ansatz~\cite{Giamarchi},
	\begin{equation}
	\begin{split}
	& K = \frac{\pi}{2\left(\pi - \arccos\Delta\right)}, \\
	& v = \frac{2v_F\left(\pi - \arccos\Delta\right)}{\pi}.
	\end{split}
	\end{equation}
	We note that for lower values of $\Delta$, corresponding to higher values of $K$, the Gaussian distribution of the charge resolved entanglement is expected to be wider, and the effect of higher orders of $\alpha$ smaller, as can be seen from Eq. (\ref{eq:tEEFlux}).

	For $\Delta = 0$ Eq. (\ref{eq:XXZf}) describes spinless noninteracting fermions, allowing an exact calculation of the entanglement. Then, the entanglement Hamiltonian $\hat{H}_A$, defined as $\rho_A = e^{-\hat{H}_A}$, is quadratic. We follow Ref.~\onlinecite{GS} (based on the method introduced in Ref.~\onlinecite{Peschel}), and obtain exact results for the entanglement entropies,
	\begin{equation}\label{eq:exact}
	S_A^{(n)} = \prod_n \left[ e^{i\alpha}(f_l)^n + (1-f_l)^n \right],
	\end{equation}
	where $f_l = 1/\left(e^{\epsilon_l} + 1\right)$, and $\epsilon_l$ are the eigenvalues of $\hat{H}_A$. $f_l$ are the eigenvalues of the subsystem correlation matrix $C_{ij} = \left\langle c_i^{\dagger} c_j \right \rangle, i,j = 1\dots L_A$, which can be obtained exactly for the noninteracting case by explicit calculation of the matrix $C$ of the full system in the momentum space $\ket{k} = \sum_{i = 1}^{N}\sin\left(\frac{\pi ik}{N+1}\right)\ket{i}$, $k=1...N$, where the basis states are eigenstates of the energy, $C_{k,k^\prime}(t) = C_{k,k^\prime}(0)e^{-i(E_{k^\prime} - E_k)t}$, for $E_k = 2J\cos\left(\frac{\pi k}{N+1}\right)$. $C(0)$ is the correlation matrix in the ground state of the decoupled system, , which is diagonal in the corresponding eigenbasis $\ket{q;L} = \sum_{i=1}^{N/2}\sin\left(\frac{\pi i q}{N/2 +1}\right)\ket{i}$ and $\ket{q;R} = \sum_{i=N/2 + 1}^{N}\sin\left(\frac{\pi (i - N/2) q}{N/2 +1}\right)\ket{i}$, for $q=1...N/2$. In this basis $Q_{q\ell, q^\prime\ell^\prime}=\delta_{qq^\prime}\delta_{\ell\ell^\prime}\Theta(-E_q)$, for $\ell,\ell^\prime=L,R$.
	\textcolor{blue}{$C = C^L\otimes C^R$ and $ C^\ell_{q, q^\prime}=\delta_{q,q^\prime}\Theta(-E_q)$.}
	%-%
	
	Results for the case in which the boundary point between $A$ and $B$ is the quench point are presented in Fig. \ref{fig:exactEE}, and for the case when the boundary point is moved away by $x = l$ from the quench point in Fig. \ref{fig:exactEEShift}. We use open boundary conditions throughout. In all cases in this study, $a$ and $w_1$ from Eq. (\ref{eq:tEEFlux}) were used as fitting parameters. 
	The entanglement is periodic in time, as predicted by Eq. (\ref{eq:transSD}). This is the result of the entanglement being carried by the quasiparticles moving in velocity $v$ as mentioned above, bumping at the ends of the system  and coming back to the other side~\cite{SD}. In Fig. \ref{fig:exactEEShift}, CFT predicts the entanglement to be constant for $\mod(vt, L) < l$ or $ \mod(vt, L) > L-l$, when these quasiparticles are allegedly outside of subsystem A. The exact results present some tails  in these time regimes, which are caused by excitations moving with velocities smaller than $v$, which are not accounted for by CFT~\cite{SD}.
	We present results both for the zeroth order, calculated from Eq. (\ref{eq:tEEFlux}) with $w_m = 0$ for $m \ne 0$, and for the first order, in which $w_1 \ne 0$.
	The zeroth order seems to be a satisfying approximation from Figs. \ref{fig:exactEE} and \ref{fig:exactEEShift}, but as can be seen in Fig. \ref{fig:exactEEAlpha}, it is insufficient for large values of $|\alpha|$.

	For nonzero values of $\Delta$, we compare the CFT prediction to numerical results obtained by the tDMRG algorithm~\cite{SCHOLLWOCK,Vidal1,Vidal2,tDMRG1,tDMRG2}, employing the QSpace tensor library~\cite{QSpace}. The extraction of the charged resolved entanglement spectrum is natural in this method, thanks to the block diagonal form of the Matrix Product State (MPS) matrices that are used to represent the state in this algorithm: in the MPS representation, one separates the state of an $N$-sites system into $N$ rank-3 tensors~\cite{SCHOLLWOCK}. One entry of the tensor is called the physical index, and ranges over the physical Hilbert space of a single site, whose dimension is 2 in our case. The other two entries connect the site to the rest of the system from left and right, and represent the Hilbert space of all the sites on the left of the site or all the sites on the right of it, truncated to the so called ``bond dimension'' see Fig. \ref{fig:MPS}. For a subsystem $A$ extending the left edge of the system to site $i$, the eigenvalues of $\rho_A$ are extracted by placing the orthogonality center~\cite{SCHOLLWOCK} at the $i$th tensor and combining its physical index and its left index, thus getting a matrix with one entry representing subsystem $A$ and the second entry representing the rest of the system to the right of $A$. One can exploit the symmetry in the system by separating these tensors into blocks corresponding to different charge sectors. From these charge sector blocks we extract the eigenvalues of each block of the RDM separately.
	
	 We used a second order Trotter approximation with a timestep of $\Delta t = 10^{-2} J^{-1}$. We set the MPS bond dimension~\cite{SCHOLLWOCK} to 1,024 and the MPS truncation error~\cite{SCHOLLWOCK} to $10^{-8}$ in all tDMRG runs for the entanglement entropy (in practice the truncation error was $~10^{-10}$ or less).
	
	In order to stay well within the CFT region, we chose $-0.5 \le \Delta \le 0.5$. Results for several values of $\Delta$ for open boundary conditions are plotted in Fig. \ref{fig:tdmrgEE}. Here too results for both $w_1 = 0$ and $w_1 \ne 0$ are presented. $w_1=0$ appears to generally be a satisfactory fit, as in the noninteracting case. We see oscillations in the numerical data not predicted by CFT: These are oscillations due to finite lattice spacing, which are absent in the CFT approximation and decrease for large $t$ (which is why they were not seen in Fig. \ref{fig:resExact}, where the system length and achievable timescales are much longer). We note that the contribution of the corresponding spatial oscillations was calculated for the ground state case of the XX model in Ref.~\onlinecite{CalEs}, but these results are not straightforward to extend to the current time-dependent case. 
			
	%\textit{Non-Integrable Systems.} Adding a next-nearest neighbors (NNN) interaction to the XXZ model turns it into a non-integrable system~\cite{Giamarchi},
	%
	%\begin{equation}\label{eq:ni}
	%\begin{aligned}
	%H = & J\sum_n\left[\sigma_n^+\sigma_{n+1}^- + \mathrm{h.c.} + \Delta\sigma_n^z\sigma_{n+1}^z \right.\\
	%& + \left. \Delta^\prime_2\sigma_n^+\sigma_{n+2}^- + \mathrm{h.c.} + \Delta_2 \sigma_n^z\sigma_{n+2}^z\right].
	%\end{aligned}
	%\end{equation}
	%Non-integrable systems are expected to thermalize after a local quantum quench as described above. The thermalization time introduces a time scale into the dynamics of the system, and it will thus no longer obey CFT. However, if the NNN interaction is weak enough, we expect CFT to describe the behavior of the system for relatively long times, which are still shorter than the thermalization time. The system can then be approximated as a Luttinger liquid, with excitation velocities and Luttinger parameter numerically calculated as in Ref.~\onlinecite{Ejima}. Fig. \ref{fig:ni} presents tDMRG results for a system with weak NNN interaction after a local quench for the full and charge-resolved vNEEs. CFT predictions become worse and worse as time progresses, but the qualitative behavior still follows CFT within the time frame accessible numerically.
	
	\subsection{\label{sec:resNeg} Entanglement Negativity}
	The numerical method for the exactly solvable XX model is developed in Ref.~\onlinecite{CGS}, and is an extension of the entanglement Hamiltonian method used for the entanglement entropy case. We rely on the result from Refs.~\onlinecite{EZ1,EZ2} that in this noninteracting case the partially transposed RDM is a sum of two Gaussian matrices: $\rho_A^{T_2} = \sum_{\sigma = \pm 1} u_\sigma \frac{\hat{O}_\sigma}{\Tr{\hat{O}_\sigma}}$. $u_\sigma$ are the coefficients $u_\sigma = \frac{1}{\sqrt{2}}e^{-\frac{\pi}{4}\sigma}$, and $\hat{O}_\sigma = e^{\sum_{ij}c_i^\dagger W^{(\sigma)}_{ij} c_j}$, where the matrices $W^{(\sigma)}_{ij}$ can be extracted from the correlation matrix $C$ defined in the
	previous subsection,
	\begin{equation}
	e^{W_\sigma} = \frac{1+G_\sigma}{1 - G_\sigma}, \quad G_\sigma = \begin{pmatrix} 
	C^{(11)} & \sigma i C^{(12)} \\
	\sigma i C^{(21)} & C^{(22)} 
	\end{pmatrix},
	\end{equation}
	where $C^{(\alpha\beta)}$ are the blocks of $C$ is the regions corresponding to subsystems $A_\alpha, A_\beta$. In Ref.~\onlinecite{CGS} it was shown that for a quadratic operator $\hat{X} = \sum_{ij} c_i^\dagger X_{ij} c_j$,
	\begin{equation}\label{eq:negsNum}
	\Tr\left[e^{i\hat{X}}\left(\rho_A^{T_2}\right)^n\right] = \sum_{\{\sigma\}} u_{\{\sigma\}} \det\left(\frac{ 1+e^X \prod_i e^{W^{(\sigma_i})} }{\prod_i (1 + e^{W^{(\sigma_i)}})}\right),
	\end{equation}
	where $u_{\{\sigma\}} = \prod_i u_{\sigma_i}$. The charge imbalance $\hat{Q}$ is a particular case of such a quadratic operator: $\hat{Q} = \sum_{i,j}c_i^\dagger (-1)^{\gamma(i)}\mathbb{I}_{ij}c_j$, where $\gamma(i) = 0$ if $i \in A_1$, and $\gamma(i) = 1$ if $i \in A_2$. Combining the equation above with Eq. (\ref{eq:rnAlpha}), which is of the form of the RHS of Eq. (\ref{eq:negsNum}), we can obtain the exact RNs for the XX model. 
	
	In this method, the only RNs that are numerically accessible are $\mathcal{N}_{A_1,A_2}^{(1)}, \mathcal{N}_{A_1,A_2}^{(2)}$, and $\mathcal{N}_{A_1,A_2}^{(3)}$, since the matrices in the denominator in Eq. (\ref{eq:negsNum}) are almost singular, and only for $n \le 3$ one can explicitly cancel the small denominator against a corresponding factor in the numerator.

	For nonzero values of $\Delta$ we used the tDMRG algorithm and extracted the spectrum of $\rho_A^{T_2}$ following the method described in Ref.~\onlinecite{RAC}. 
	A naive method for extracting the negativity of an MPS goes as follows: We contract all of the tensors representing one subsystem $A_i$ into one big tensor, and combine all the physical indices to one physical index of degree $2^{|A_i|}$. We then explicitly create $\rho_A^{T_2}$ by taking two copies of each rank-3 tensor, tracing out the environment, and applying a partial transpose to the physical indices of subsystem $A_2$, see Fig. \ref{fig:dmrgNeg}. This partial transpose guarantees that the block form of the matrix, formerly corresponding to the charge on both subsystems, will now correspond to the charge imbalance, as illustrated in Fig. \ref{fig:dmrgneg3}. The extraction of the tensor in this method for an MPS bond dimension $D$ is $D^8$. In Ref.~\onlinecite{RAC} a more sophisticated way of constructing an equivalent tensor is presented in complexity $D^6$, which has the same charge imbalance block structure. We use the latter method in our calculations.
	
	In this method all negativities are accessible, and we can use the block structure just explained for extracting the charge imbalance resolved negativities. However, since the dependence of the runtime on the bond dimension is stronger for the extraction of the negativity as compared to the entropy, we reduced the bond dimension to 256, leading to a truncation error of $\sim10^{-5}$.

	By definition, $\mathcal{N}_{A_1,A_2}^{(1)}(Q)$ is simply the probability distribution of $Q$. However, it is worthwhile to fit it to the CFT prediction. The reason is that for $n = 1$ we do not apply any twist fields. In this case, our vertex operator 6-point function is just an ordinary free-boson 6-point function, for which the nonuniversal function $\mathcal{F}$ should be equal to unity~\cite{BYB}. Fig. \ref{fig:negR1} presents the fit to Eq. (\ref{eq:tRNFluxWpm}) for the XX model for a very large system using the transformation (\ref{eq:transCC}), and for the interacting case (using tDMRG) for a finite system using the transformation (\ref{eq:transSD}). The fits are seen to work well in the current $n = 1$ case.

\begin{figure}[t]
		\centering
		\subfloat[\label{fig:neg3d0}]{
			\includegraphics[width=1\linewidth]{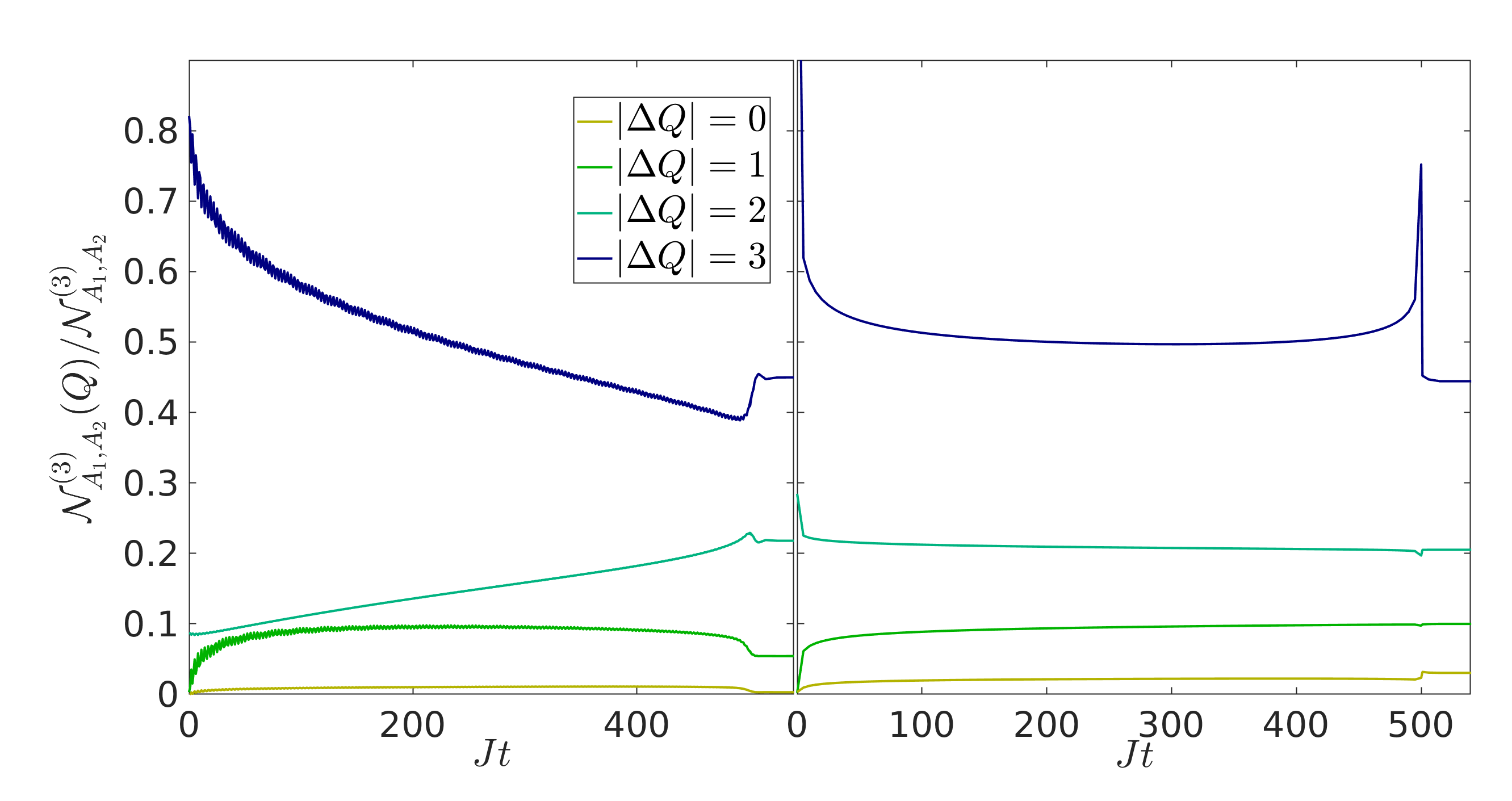}	
		} \\		
		\subfloat[\label{fig:neg3d-05}]{
			\includegraphics[width=1\linewidth]{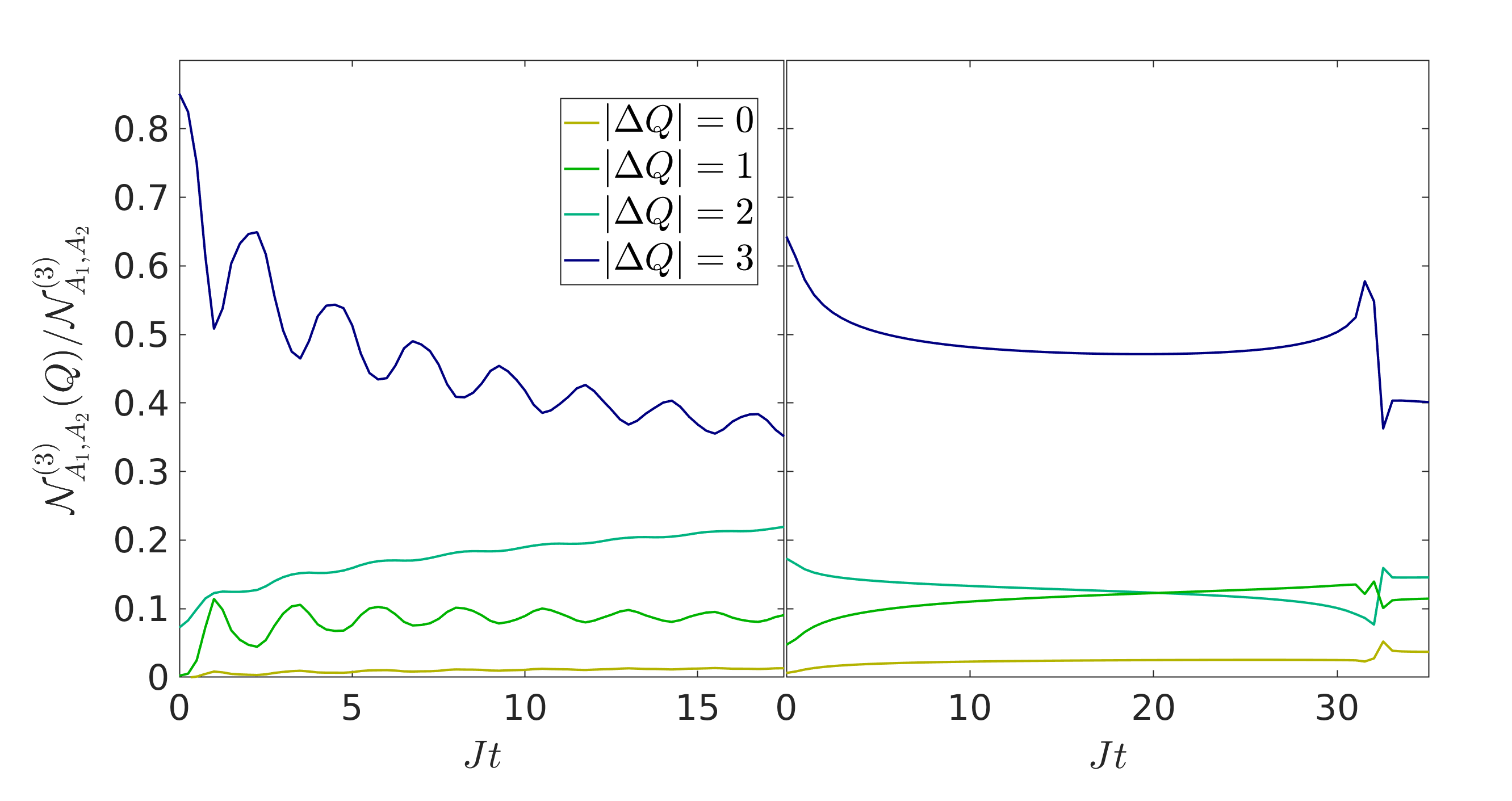}	
		}
		\caption{Comparison between numerical results (left) and CFT predictions (right) for the third imbalance resolved RN. In (a) the XX model results are presented for a large system ($\frac{L}{a} = 10,000$), for $\frac{L_{A_i}}{a} = 1000$. (b) Results for the XXZ model with $\Delta = -0.5$, for $\frac{L}{a} = 256$ and $l = \frac{L_{A_i}}{a} = 64$.}
		\label{fig:negR3}
	\end{figure}

	The second RN is by definition simply the second RE, or purity, of $A = A_1 \cup A_2$, as mentioned in Ref.~\onlinecite{CCT}. We thus skip to the first non-trivial RN, $\mathcal{N}_{A_1,A_2}^{(3)}$. Fig. \ref{fig:negR3} presents comparisons of the CFT prediction to the exact ($\Delta=0$) and tDMRG ($\Delta \ne 0$) results.
	Here things get complicated: It looks like the function $\mathcal{F}$ modifies some of the qualitative behavior of the prediction as well.  We see that while the overall size and trends of the charge-resolved negativities is generally correct for $vt \ll l$, the qualitative behavior is not always fully reproduced. We are left with the conclusion that the effect of the non-universal contribution for the vertex operators 6-point function (imbalance-resolved RNs) has a more pronounced effect than it does for the twist fields 6-point function (total RNs), similarly to the static case~\cite{WCR}.

	\section{\label{sec:conclusion} Conclusions and Future Outlook}
	We expanded the understanding of the charge-resolved entanglement entropies, defined in Ref.~\onlinecite{GS}, and	 charge-imbalance-resolved entanglement negativities, defined in Ref.~\onlinecite{CGS}, by studying for the first time the time-dependence of these quantities following a local quench. We started by analytically calculating the time dependent charge resolved entanglement for 1+1D CFT: We added flux-like vertex operators to the time dependent replica trick following Ref.~\onlinecite{GS}. We then compared the prediction to exact results for the noninteracting XX model, as well as tDMRG results for the critical range of the XXZ model. While the total entanglement scales as the log of the protocol-dependent effective length $L_\mathrm{eff}(t)$ (Eq. (\ref{eq:tee})), we find that the contribution comes from  $\propto\sqrt{\log(L_\mathrm{eff}(t))}$ significant charge sectors, each contributing an amount proportional to $\sqrt{\log(L_\mathrm{eff}(t))}$ to the entanglement (Eqs. (\ref{eq:gsRes}, \ref{eq:tEEFlux})). For example, this could be used to understand the number of states one needs to keep in each conserved number block in tDMRG~\cite{dmrgScale}.
	
	The dependence of the flux-resolved entanglement on the flux appears to be more complicated for smaller systems (whether the entire system is small in size, or just time is short, making the effective length small), and we needed to amend our expression by expanding the vertex operator as in Eq. (\ref{eq:wpm}). However, the CFT predictions for the vNEE agrees nicely with the numerical results for Luttinger liquids, both with and without taking the full expansion for the vertex operator.
	
	For the charge imbalance resolved negativity, the time-dependent calculation forced the use of boundary CFT. In this case, high-order correlation functions appear, which are non-universal. General CFT arguments are sufficient for obtaining the qualitative behavior, but the calculation of the full nonuniversal confromal block remains an interesting question for the future.
	
	Our results pave the way towards studying additional models, such as 1+1D conformal systems with non-abelian symmetries~\cite{GS} or systems undergoing global quenches (based on the the appropriate conformal transformation from Ref.~\onlinecite{CC}). It will also be interesting to study the behavior of charge resolved entanglement in non-CFT systems (e.g., topological ones ~\cite{Kitaev06,Levin06,Shtengel}) or conformal systems of higher dimension. One possible motivation is that the charge-resolved entanglement measures are much more numerically accessible for high values of $\Delta N_A$ or $\Delta Q$ --- these charge or charge imbalance sectors carry less entanglement and hence their corresponding blocks in $\rho_A, \rho_A^{T_2}$ respectively are smaller and easier to diagonalize~\cite{LR}. Thus, if the expected $\Delta N_A$ or $\Delta Q$ is known, one can use calculations in these smaller sectors to characterize all sectors.
	
	\begin{acknowledgments}
		We would like to thank G. Cohen, E. Cornfeld, E. Grosfeld, and E. Sela for useful discussions.
		Support by the Israel Science Foundation (Grant No. 227/15), the German Israeli Foundation (Grant No. I-1259-303.10), the US-Israel Binational Science Foundation (Grant No. 2016224), and the Israel Ministry of Science and Technology (Contract No. 3-12419) is gratefully acknowledged.
	\end{acknowledgments}

	\appendix
	\section{Full Formulas for the Symmetry-Resolved Entanglement Entropies}\label{app:entropy}
	The final formula used for fitting the numerical results is a combination of Eqs. (\ref{eq:transSD}), (\ref{eq:fourier}) and (\ref{eq:tEEFlux}) from the main text. These include the effective length $L_\mathrm{eff}(t)$.
	For the entanglement entropy, we treated two cases. In the first case, where $L_A=L_B=L/2$ and the quench point is the boundary point between the two subsystems, we get from Eq. (\ref{eq:transSD})
	\begin{equation}
	L_\mathrm{eff} = \frac{L}{\pi}\sqrt{\sinh\left(\frac{\pi}{L}\left(\epsilon + iv t\right)\right)\sinh\left(\frac{\pi}{L}\left(\epsilon - i v t\right)\right)}.
	\end{equation}
	When the quench point is in the middle of the full system ($x = L/2$), but the boundary point between the two subsystems is shifted from the quench point by $l$, so that $L_A=L/2-l$, $L_B=L/2+l$, we obtain, using the conformal transformation (\ref{eq:transSD}) again but now adding the twist field at $z = l + i\tau$,
\begin{widetext}
			\begin{equation}
		L_\mathrm{eff} = \frac{L}{\pi}\sqrt{
			\left(\cos^2\frac{\pi(vt + i\epsilon)}{L} - \cos^2\frac{\pi l}{L}\right) \times \left(\cos\frac{2\pi l}{L}\cosh\frac{2\pi \epsilon}{L} - \cos\frac{2\pi vt}{L} + \right. \left. 2\left|\cos^2\frac{\pi(vt - i\epsilon)}{L} - \cos^2\frac{\pi l}{L}\right|\right)
		}\quad.
		\end{equation}

	Both terms can be calculated straightforwardly by performing the conformal transformation (\ref{eq:transSD}) and combining it with Eq. (29) in Ref.~\onlinecite{CC}.
	
	We now substitute $L_\mathrm{eff}$ into Eqs. (\ref{eq:scalingD}) and (\ref{eq:wpm}), and get:
	\begin{equation} \label{eq:fluxRenyiFinal}
	\begin{aligned}
	S_A^{(n)}\left(\alpha\right) =  (L_{\mathrm{eff}}/a)^{-\left(\frac{c}{12}\left(n - n^{-1}\right) +  \frac{K}{n}\left(\frac{\alpha}{2\pi}\right)^2\right)} +  w_1^2(L_{\mathrm{eff}}/a)^{-\left(\frac{c}{12}\left(n - n^{-1}\right) + \frac{K}{n}\left(\frac{\alpha + 2\pi}{2\pi}\right)^2\right)} +  w_1^2(L_{\mathrm{eff}}/a)^{-\left(\frac{c}{12}\left(n - n^{-1}\right) +  \frac{K}{n}\left(\frac{\alpha - 2\pi}{2\pi}\right)^2\right)}.
	\end{aligned}
	\end{equation}
	We can now perform a derivation by $n$ in order to get the flux-resolved vNEE:
	\begin{equation}
	\begin{aligned}
	S_A(\alpha) = & -\frac{1}{4 \pi ^2}\log (L_\mathrm{eff}/a) (L_\mathrm{eff}/a)^{\frac{\alpha  (\alpha -4 \pi ) K} {4 \pi ^2}}  \left(w_1
	\left(\alpha ^2 K+4 \pi  \alpha  K+4 \pi ^2 (2 c+K)\right)  (L_\mathrm{eff}/a)^{\frac{2 \alpha 
			K}{\pi }+K} + \right. \\& \left.
	\left(\alpha ^2 K+8 \pi ^2 c\right) (L_\mathrm{eff}/a)^{\frac{\alpha  K}{\pi
	}}+ 
	w_1 L^K \left(\alpha ^2 K-4 \pi  \alpha  K+4 \pi ^2 (2
	c+K)\right)\right),
	\end{aligned}
	\end{equation}
	following a Fourier transform (Eq. (\ref{eq:fourier})) which will gives us the charge-resolved vNEE:
	\begin{equation}
	\begin{aligned}
	S_A(N_A) = & e^{-2 i \pi  N_A} \left(i \sqrt{\pi } \left(-w_1+e^{2 i \pi  N_A}\right)
	e^{\frac{\pi ^2 N_A^2}{K \log (L_\mathrm{eff}(t)/a)}}  \times\right. \\& \left. \left(K \log (L_\mathrm{eff}(t)/a) (1-4 c \log (L))+2 \pi ^2
	N_A^2\right)  
	\text{erf}\left(\frac{2 \pi  N_A-i K \log (L_\mathrm{eff}(t)/a)}{2 \sqrt{K} \sqrt{\log
			(L_\mathrm{eff}(t)/a)}}\right)+ \right. \\ & \left.
	i \sqrt{\pi } \left(-1+e^{2 i \pi  N_A} w_1\right) e^{\pi  N_A
		\left(\frac{\pi  N_A}{K \log (L_\mathrm{eff}(t)/a)}+2 i\right)} \times\right.\\&\left.
	\left(K \log (L_\mathrm{eff}(t)/a) (1-4 c \log (L_\mathrm{eff}(t)/a))+2
	\pi ^2 N_A^2\right) 
	\text{erf}\left(\frac{2 \pi  N_A+i K \log (L_\mathrm{eff}(t)/a)}{2 \sqrt{K}
		\sqrt{\log (L_\mathrm{eff}(t)/a)}}\right)+ \right.\\&\left.
	i \sqrt{\pi } w_1 e^{\frac{\pi ^2 N_A^2}{K \log
			(L_\mathrm{eff}(t)/a)}} 
	\left(K \log (L_\mathrm{eff}(t)/a) (1-4 c \log (L_\mathrm{eff}(t)/a))+2 \pi ^2 N_A^2\right) \times\right.\\&\left.
	\left(\text{erf}\left(\frac{2 \pi  N_A-3 i K \log (L_\mathrm{eff}(t)/a)}{2 \sqrt{K} \sqrt{\log
			(L_\mathrm{eff}(t)/a)}}\right)-  
	e^{4 i \pi  N_A} \text{erf}\left(\frac{2 \pi  N_A+3 i K \log (L_\mathrm{eff}(t)/a)}{2
		\sqrt{K} \sqrt{\log (L_\mathrm{eff}(t)/a)}}\right)\right)+ \right.\\&\left.
	2 \sqrt{K} e^{2 i \pi  N_A} (L_\mathrm{eff}(t)/a)^{K/4}
	\sqrt{\log (L_\mathrm{eff}(t)/a)} \times
	\right.\\&\left.
	(K (w_1-1) \log (L_\mathrm{eff}(t)/a) \cos (\pi  N_A)- 
	2 \pi  N_A (w_1+1)
	\sin (\pi  N_A))- \right.\\&\left.
	2 \sqrt{K} e^{2 i \pi  N_A} w_1 (L_\mathrm{eff}(t)/a)^{9 K/4} \sqrt{\log (L_\mathrm{eff}(t)/a)} \times\right.\\&\left.
	(3
	K \log (L_\mathrm{eff}(t)/a) \cos (\pi  N_A)+2 \pi  N_A \sin (\pi  N_A))\right) / \left(
	4 K^{3/2} \log
	^{\frac{3}{2}}(L_\mathrm{eff}(t)/a) \right),
	\end{aligned}
	\end{equation}
where $\text{erf} = \frac{2}{\sqrt{\pi}}\int_0^z e^{-t^2} dt$ is the error function\cite{AS}.
\end{widetext}

\section{Final Formulas for the Entanglement Negativity}\label{app:neg}

The final formula for the flux resolved entanglement negativities, Eq. (\ref{eq:tRNFluxWpm}), involves the expressions $|\omega^\prime_i - \omega^\prime_j|$ defined in the main text. These expressions for the case of two adjoint subsystems of size $l$ (for an infinite total system) at time $t$ after a local quench were derived in Ref.~\onlinecite{WCR}, and are reproduced here for completeness:
	\newline
	\newline
	\newline
	\newline

\begin{equation}
\begin{aligned}
|\omega_1 - \omega_2| = \begin{cases}
\sqrt{\frac{l}{l+vt}} , \quad vt < l, \\
\sqrt{\frac{l}{vt(vt+l)}} , \quad vt > l,
\end{cases} \\
|\omega_1 - \tilde{\omega}_2| = \begin{cases}
\sqrt{\frac{l}{l-vt}} , \quad vt < l, \\
\frac{2}{\epsilon}\sqrt{vt(vt - l)} , \quad vt > l,
\end{cases} \\
|\omega_1 - \omega_3| = \begin{cases}
\frac{2}{\epsilon}\sqrt{l^2 - (vt)^2} , \quad vt < l, \\
2l\sqrt{\frac{1}{(vt)^2-l^2}} , \quad t > l,
\end{cases}\\
|\omega_1 - \tilde{\omega}_3| = \begin{cases}
\frac{2}{\epsilon}\sqrt{l^2 - (vt)^2}  , \quad vt < l, \\
\frac{2}{\epsilon}\sqrt{(vt)^2 - l^2}  , \quad vt > l,
\end{cases} \\
|\omega_2 - \omega_3| = \begin{cases}
\frac{2}{\epsilon}\sqrt{l(l-vt)}  , \quad t < l, \\
\sqrt{\frac{l^2}{vt(vt-l)}} , \quad vt > l,
\end{cases} \\
|\omega_2 - \tilde{\omega}_3| = \begin{cases}
\frac{2}{\epsilon}\sqrt{l(l+vt)}   , \quad vt < l, \\
\frac{2}{\epsilon}\sqrt{vt(l+vt)}   , \quad vt > l.
\end{cases} \\
\end{aligned}
\end{equation}

%\bibliographystyle{apsrev4-1}
%\bibliography{charge_resolved_entanglement_quench}
%merlin.mbs apsrev4-1.bst 2010-07-25 4.21a (PWD, AO, DPC) hacked
%Control: key (0)
%Control: author (72) initials jnrlst
%Control: editor formatted (1) identically to author
%Control: production of article title (-1) disabled
%Control: page (0) single
%Control: year (1) truncated
%Control: production of eprint (0) enabled
%

\end{document}